\documentclass[conference]{IEEEtran}
\IEEEoverridecommandlockouts
\usepackage{hyperref}
\usepackage{multirow}
\usepackage{graphicx}
\usepackage{graphics}
\usepackage{subcaption}
\usepackage{amsmath}
\usepackage{algorithm}
\usepackage{algorithmic}
\usepackage{wrapfig}
\usepackage{caption}
\usepackage{float}
\usepackage[normalem]{ulem}
\useunder{\uline}{\ul}{}
\usepackage{bbding}
\usepackage{pifont}
\usepackage{amssymb}

\def\BibTeX{{\rm B\kern-.05em{\sc i\kern-.025em b}\kern-.08em
    T\kern-.1667em\lower.7ex\hbox{E}\kern-.125emX}}
\begin{document}

\title{Distributional Domain-Invariant Preference Matching for Cross-Domain Recommendation}

\author{\IEEEauthorblockN{Jing Du*}
\IEEEauthorblockA{
\textit{School of Computer Science and Engineering}\\
\textit{The University of New South Wales} \\
jing.du2@unsw.edu.au}
\and
\IEEEauthorblockN{Zesheng Ye*}
\IEEEauthorblockA{
\textit{School of Computer Science and Engineering}\\
\textit{The University of New South Wales} \\
zesheng.ye@unsw.edu.au}
\and
\IEEEauthorblockN{Bin Guo}
\IEEEauthorblockA{
\textit{School of Computer Science} \\
\textit{Northwestern Polytechnical University}\\
guobin.keio@gmail.com}
\and
\IEEEauthorblockN{Zhiwen Yu}
\IEEEauthorblockA{
\textit{School of Computer Science} \\
\textit{Northwestern Polytechnical University}\\
zhiweny@gmail.com}
\and
\IEEEauthorblockN{Lina Yao$\dagger$}
\IEEEauthorblockA{
\textit{CSIRO's Data 61 and} \\
\textit{The University of New South Wales} \\
lina.yao@data61.csiro.au}
\thanks{* Jing Du and Zesheng Ye are co-first authors with equal contributions.}
\thanks{$\dagger$ Lina Yao is also affiliated with Macquarie University.}
}

\maketitle

\begin{abstract}
Learning accurate cross-domain preference mappings in the absence of overlapped users/items has presented a persistent challenge in Non-overlapping Cross-domain Recommendation (NOCDR).
Despite the efforts made in previous studies to address NOCDR, several limitations still exist.
Specifically,
1) while some approaches substitute overlapping users/items with overlapping behaviors, they cannot handle NOCDR scenarios where such {\it auxiliary information} is unavailable;
2) often, cross-domain preference mapping is modeled by learning {\it deterministic explicit representation} matchings between sampled users in two domains. 
However, this can be biased due to individual preferences and thus fails to incorporate preference continuity and universality of the general population.
In light of this, we assume that despite the scattered nature of user behaviors, there exists a consistent latent preference distribution shared among common people.
Modeling such distributions further allows us to capture the continuity in user behaviors within each domain and discover preference invariance across domains.
To this end, we propose a \textbf{D}istributional domain-invariant \textbf{P}reference \textbf{M}atching method for non-overlapping \textbf{C}ross-\textbf{D}omain \textbf{R}ecommendation (\textbf{DPMCDR}).
For each domain, we hierarchically approximate a posterior of domain-level preference distribution with empirical evidence derived from user-item interactions.
Next, we aim to build {\it distributional implicit} matchings between the domain-level preferences of two domains.
This process involves mapping them to a shared latent space and seeking a consensus on domain-invariant preference by minimizing the distance between their distributional representations therein.
In this way, we can identify the alignment of two non-overlapping domains if they exhibit similar patterns of domain-invariant preference.
Experiments on real-world datasets demonstrate that DPMCDR outperforms the state-of-the-art approaches with a range of evaluation metrics.
\end{abstract}


\section{Introduction}

Cross-domain recommendation(CDR) is widely considered an effective approach to tackle the long-standing data scarcity issue in recommender systems~\cite{khan2017cross} by transferring knowledge of users/items available in one domain to another~\cite{kang2019semi,zhu2021transfer}.
Collaborative Filtering~(CF) has emerged as a widely explored approach, where Matrix Factorization~(MF)\cite{singh2008relational, hu2013personalized} and Neural Networks~(NN)\cite{cao2022disencdr, zhu2022personalized} have been actively employed in the context of CDR.
MF-based methods aim to discover user similarities from observed user-item interactions to facilitate knowledge transfer.
For instance, Hu et al.~\cite{hu2013personalized} propose to capture cross-domain factors by horizontally connecting the interaction matrices of different domains.
However, MF-based methods are highly dependent on observation availability and thus tend to perform poorly in cold-start settings.
To overcome it, recent NN-based methods have adopted Embedding-and-Mapping~\cite{man2017cross} to adapt the information across different domains effectively. 
As an example, PTUPCDR~\cite{zhu2022personalized} use a meta-network to generate personalized bridge functions based on user representations to transfer personalized preferences. 
Nevertheless, these approaches necessitate overlapping users/items to develop reliable representations and capture domain correlations~\cite{cantador2015cross}.
Their performance would be compromised in the absence of overlapping users/items, leading to what is known as the Non-Overlapping Cross-Domain Recommendation(NOCDR) problem.
Prior NOCDR studies have primarily explored auxiliary user/item profiles or behaviors as substitutes for unavailable users/items~\cite{zang2021survey}.
In this case, Liu et al.~\cite{liu2022collaborative} connect non-overlapping users through the overlapped review attributes of users.
Likewise, users exhibiting similar rating patterns can be linked based on social behaviors~\cite{shu2018crossfire}.

\begin{figure}[]
    \centering
    \includegraphics[width=\linewidth]{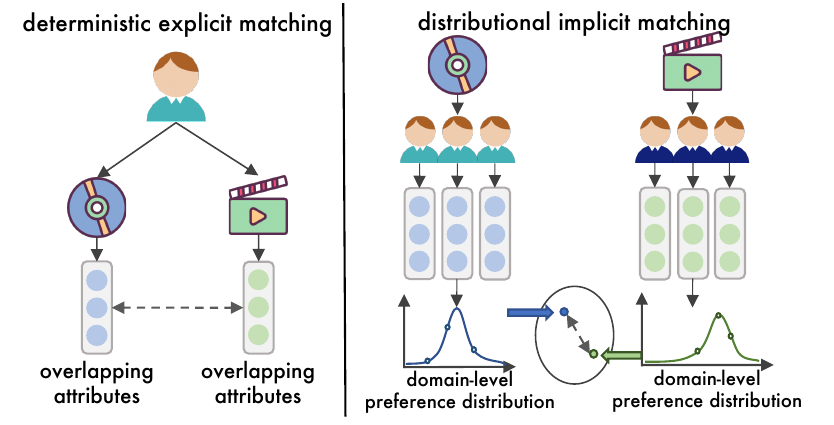}
    \caption{Difference between matching with {\it deterministic explicit matching} (left) and our {\it distributional implicit} matching (right).}
    \label{fig: example}
\end{figure}

\subsubsection*{Limitations of Previous Works}
Having said that, two central limitations remain in applying existing CDR approaches to NOCDR settings.
First, existing methods primarily focus on constructing {\it deterministic} cross-domain mappings between sampled individuals, regardless of OCDR or NOCDR, shown in Fig.~(\ref{fig: example}).
Specifically, the OCDR method PTUPCDR matches the representations of the same user present in both domains;
while in~\cite{liu2022collaborative}, the individual review attributes are aligned between domains for NOCDR.
Such {\it deterministic explicit} matching, however, is susceptible to {\it individual biases} introduced by sampling.
It can only capture the differences between individuals, thus struggling to uncover common preferences within the population~\cite{wang2021low}.
On the other hand, 
NOCDR methods that integrate auxiliary behaviors essentially pass the needs for overlapping information on to additional data from users and items, such as ratings, tags, and reviews~\cite{shu2018crossfire, zhang2019cross, liu2022collaborative}.
In~\cite{he2018robust}, domain knowledge is embedded into a cluster-level full rating matrix, allowing for the transfer of multiple rating patterns within the matrix from the source to the target domain.
In this context, {\it explicit} mapping can still be learned to some extent.
Even so, their limitations become evident in more challenging NOCDR scenarios where no auxiliary information is available.
The lack of overlapping information renders those methods that rely on {\it explicit} mapping relations ineffective, thus unsuitable for addressing NOCDR.


\subsubsection*{Research Motivation}
Still, recommendation services are intended for the general public and should be designed to follow a consistent pattern of underlying user preferences that govern user behaviors, regardless of specific domains~\cite{zang2021survey}.
Intuitively, NOCDR can take advantage of these intrinsic domain-invariant preferences.
However, related exploration remains under-investigated; the potential benefits are not fully realized by previous NOCDR methods~\cite{wang2021low}.
In light of this, we assume that there exists a continuous prior distribution of domain-level preferences, which implicitly describes the general preference pattern within each domain.
By approximating the posterior distribution, we can obtain latent cross-domain invariant preferences through alignment with different domains.
Given the limitations of {\it explicit} cross-domain matching in the absence of exact individual mapping relationships~\cite{man2017cross, cao2022cross, cao2022disencdr}, we propose matching the predictive distributions of these domain-invariant preferences instead.
The key motivation is to ensure that domain-invariant preferences, which capture the inherent patterns of users, remain highly similar irrespective of the specific domain they originate from.
To achieve this, we first use hierarchical probabilistic modeling to derive domain-level user preferences by leveraging groups of user representations from each domain, whilst considering their inner correlations.
The domain-level preferences further parameterize a predictive distribution of domain-invariant preference from both source and target domain perspectives.
Transferring these between domains can be seen as sharing the underlying universality with others.
In this way, we identify a cross-domain invariant preference by aligning two predictive distributions derived from random groups of users, referred to as {\it distributional implicit} matching.
Intuitively, this implies reaching a consensus across domains on such a cross-domain invariant preference.
Moreover, {\it implicit} distributional matching could benefit from reduced sampling bias with monte-carlo methods~\cite{harrison2010introduction}.

\subsubsection*{Developed Method}
We propose a \textbf{D}istributional Domain-invariant \textbf{P}reference \textbf{M}atching approach for non-overlapping \textbf{C}ross-\textbf{D}omain \textbf{R}ecommendation, named \textbf{DPMCDR}.
We consider the user population's preference as a continuous prior distribution across different domains.
To approximate the posterior within each domain, we leverage groups of users and derive domain-level latent representations and cross-domain invariant preferences.
By considering preferences as a collective whole, one can draw correlations between users and reduce individual bias.
Towards this, we first compute deterministic user/item representations with \textit{Deterministic Graph Encoders} from observed user-item interactions.
Then, we design {\it Stochastic Latent Preference Identifier} for posterior approximation of domain-level preference with random groups of latent user representations.
This is followed by a {\it Distributional Preference Matching} that parameterizes a predictive distribution of cross-domain invariant preference in a shared latent space for both domains.
To ensure consistency of cross-domain invariant preference, we construct a bi-directional transfer path by minimizing the Jenson-Shannon divergence between them.
In addition, we implement \textit{User-specific Optimizers} and \textit{Domain-specific Optimizers} informed by variational Information Bottleneck~(VIB)\cite{alemi2016deep} to constrain the latent representations and domain-level preference in terms of stronger generalization~\cite{tishby2015deep} for improved prediction performance.

\subsubsection*{Contributions}
Briefly, our contributions are as follows:
\begin{itemize}
    \item 
    We propose a cross-domain invariant preference matching approach for NOCDR,
    which models the {\it domain-level} preferences as continuous distributions and 
    yields a cross-domain invariant preference aligned across domains, in view of distributional {\it implicit} preference matching.
    
    \item
    Specifically, we approximate the continuous \textit{domain-level} preferences with groups of random users in both domains and parameterize two distributions of \textit{cross-domain invariant} preferences therefrom.
    With latent correlations among users incorporated, two distributions align different interpretations of the preference commonality.
    
    \item
    In addition, we present two VIB-informed optimizers that constrain latent representations to be maximally compact and informative about the user-item interactions.
    The {\it user-specific optimizer} helps to obtain robust user representations within each domain, while the {\it domain-specific optimizer} facilitates the generalization of domain-level preference in another domain.
\end{itemize}

\section{Related work}
We categorize a CDR model as either OCDR or NOCDR based on the presence or absence of overlapping users/items.
\subsection{Overlapping Cross-Domain Recommendation}
The OCDR scenario involves partial or complete user-item interactions with both domains.
In this context, OCDR models basically strive to capture the representations of overlap users from observed interactions and align/transfer them across domains \cite{man2017cross} with different methodologies, such as Matrix Factorization(MF)-based methods\cite{singh2008relational, ma2008sorec, hu2013personalized, yang2017multi} and Neural Network(NN)-based methods~\cite{bi2020dcdir, bi2020heterogeneous, zhu2022personalized, cao2022disencdr, cao2022cross}.

For MF-based methods, 
CMF~\cite{singh2008relational} share user parameters across domains to transfer knowledge.
Hu et al.~\cite{hu2013personalized} horizontally connect the matrices of different domains to obtain potential users and item factors for prediction.
Moreover, MPF~\cite{yang2017multi} analyze user behaviors on different websites and applies the probabilistic matrix factorization to capture cross-site user preferences for knowledge transfer.
The cold-start problems, however, may prevent the use of MF-based methods, leading to the surge of NN-based methods, which have demonstrated improved capacity under cold-start settings.
Man et al.~\cite{man2017cross} define the Embedding-and-Mapping paradigm and inspire a series of OCDR models.
Following pre-training of user and item embeddings, this paradigm uses overlapping correlations between users/items to learn a mapping function.
For example, DCDIR~\cite{bi2020dcdir} and HCDIR~\cite{bi2020heterogeneous} construct heterogeneous information networks and  leverage the rich information available in the user-item interactions.
Zhu et al.~\cite{zhu2022personalized} propose a meta-network to generate personalized user preference transfer bridges based on user representations, enabling the transfer of personalized preferences.
VDEA~\cite{liu2022exploiting} optimize the variational lower bound of Mixture-Of-Gaussian and align the embedding distribution of overlapping and non-overlapping users at the user level.
More recently, DisenCDR~\cite{cao2022disencdr} and CDRIB~\cite{cao2022cross} harness mutual information to filter informative user/item representations for knowledge transfer.
In summary, OCDR methods urge the existence of overlapping users/items to transfer knowledge between domains.
Their performances will be significantly reduced when faced with NOCDR scenarios.

\subsection{Non-overlapping Cross-Domain Recommendation}
A NOCDR setting suggests that no user has interactions in both domains simultaneously.
Solutions involve finding {\it explicit} correlations of user behaviors as substitutes for overlapping users/items\cite{wang2021low}. 
Yang et al.\cite{yang2015graph} connect different items with the same tags to facilitate semantic matching of items based on shared tags.
Wang et al.\cite{wang2019solving} capture user sequential behaviors by jointly embedding the rated items into a unified space.
Liu et al.\cite{liu2022collaborative} merge latent embeddings of reviews and attributions across domains and reduce the domain discrepancy with attribution alignment.
However, these approaches rely heavily on auxiliary information and overlapping behaviors for knowledge matching.
In contrast, a series of works known as codebook-based methods develop methods around cluster-level rating patterns compressed into a codebook.
Li et al.\cite{li2009can} compress the rating matrix into a compact representation in the source domain, and reconstruct the target matrix by expanding this codebook.
Subsequent works~\cite{moreno2012talmud, gao2013cross, hamilton2017inductive, zhang2018cross} direct their efforts on cluster-wise correspondence, placing a strong assumption of cluster correspondence and relying on discrete {\it explicit} matching to transfer knowledge.
A more recent study~\cite{li2022gromov} assumes that the exact rating patterns between domains can be aligned with iterative optimization.
These deterministic methods, relying on explicit relations or sampled users, may result in \textit{individual bias} among users who behave similarly in source and target domains but can be inconsistently between discrete groups.
In contrast, our method assumes the continuous distribution of underlying preferences and models the cross-domain invariance from the {\it implicit} distributional perspective, eventually reducing the {\it individual bias}.
Noticeably, our method is effective when only user-item interactions exist, without the need for auxiliary information as required by previous studies.

\begin{figure}
    \centering
    \includegraphics[width=\linewidth]{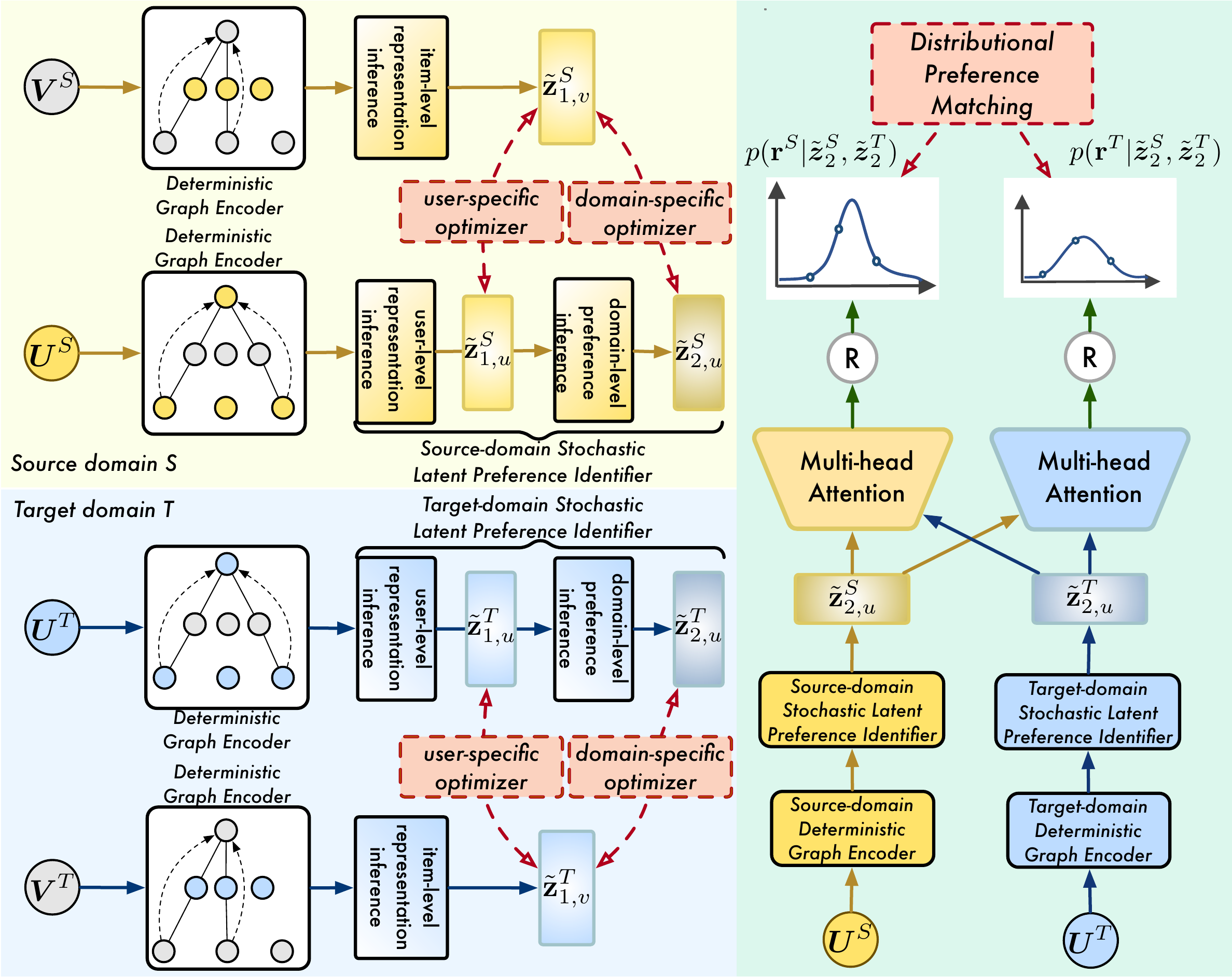}
    \caption{Framework of DPMCDR, containing {\it Deterministic Graph Encoders} and {\it Stochastic Latent Preference Identifiers} as the forward modules, while the backward information is conveyed by {\it user-specific optimizers}, {\it domain-specific optimizers} and {\it Distributional Preference Matching}.
    }
    \label{fig: framework}
    \vskip -0.15in
\end{figure}

\section{Problem Definition}
We target a NOCDR scenario\footnote{We discuss NOCDR to demonstrate the method capability in an extremely challenging setting. We note that our method can handle OCDR as well.} where only user-item interactions in either domain are available.
For each domain, we have two node sets: a user set $\mathcal{U}=\{u_i\}_{i=1:|\mathcal{U}|}$ and an item set $\mathcal{V}=\{v_i\}_{i=1:|\mathcal{V}|}$.
The user-item interactions constitute the edge set $\mathcal{R}=\left< r_{u_i,v_j} \right>$ with $u_i \in \mathcal{U}$ and $v_j \in \mathcal{V}$, indicating the interaction of user $u_i$ and item $v_j$.
If user $u_i$ has interacted with item $v_j$, then $r_{u_i,v_j}=1$, else $r_{u_i,v_j}=0$.

Given the source domain $S$ and target domain $T$, we generate domain-specific interaction bipartite graphs $\mathcal{G}^S= \left<\mathcal{U}^S, \mathcal{V}^S, \mathcal{R}^S \right>$ and $\mathcal{G}^T=\left<\mathcal{U}^T, \mathcal{V}^T, \mathcal{R}^T \right>$, with disjoint node sets $\mathcal{U}^S \cap \mathcal{U}^T = \varnothing$, and $\mathcal{V}^S \cap \mathcal{V}^T= \varnothing$,
i.e., $\mathcal{S}$ and $\mathcal{T}$ share no users and items.
We further denote the edge sets $\mathcal{R}^S$ and $\mathcal{R}^T$ by two adjacency matrices $\boldsymbol{A}^S \in \mathbb{R}^{|\mathcal{U}^S| \times |\mathcal{V}^S|}$ and $\boldsymbol{A}^T \in \mathbb{R}^{|\mathcal{U}^T| \times |\mathcal{V}^T|}$, where each element $a_{i,j}$ refers to the interaction status from user $u_i$ to item $v_j$.

In this section, we focus on a cold-start setting.
Given a new user with no interactions in either domain, we predict the item to be interacted with, by approaching {\it distributional preference matching} between the source and the target domain.

\section{Methodology}
\subsection{Overview}
Fig.~(\ref{fig: framework}) provides an overview of the DPMCDR framework. 
DPMCDR utilizes {\it Deterministic Graph Encoders} to aggregate homogeneous neighbors of each node from user-item interactions, deriving user/item-specific representations in each domain. 
Then, {\it Stochastic Latent Preference Identifiers} approximate the posterior of latent user/item representations and domain-level preferences to capture preference continuity and commonality in user behaviors, presumed to be stable regardless of domains.

Concerning cross-domain alignment, DPMCDR includes the {\it Distributional Preference Matching}, which parameterizes a Gussian-based cross-domain invariant preference distribution using random groups of users from both domains.
Two outcomes result from this: domain-level preference distributions for both source and target domains, reflecting cross-domain invariance from the respective perspectives of each domain.
Model optimization in DPMCDR takes place through \textit{user-specific optimizer}, \textit{domain-specific optimizer}, and \textit{distributional preference matching optimizer}.
While the first two optimizers guide each representation with predictive information about NOCDR task, the last matching objective ensures invariant preferences are aligned across domains.

\subsection{Deterministic Graph Encoder}
\label{section: encoder}
We first obtain initial user and item representations from the user-item interaction graph $\mathcal{G}$ in each domain.
Using the graph convolutional network~(GCN)~\cite{kipf2016semi}, we can compute user- and item-specific representations from their homogeneous neighbors.
Unlike previous GCN-based works that directly aggregate one-hop\footnote{In the interaction bipartite graph $\mathcal{G}$, a user can have no one-hop neighbors other than an item. The homogeneous nodes can thus only be indirectly linked.}
neighbors~\cite{du2023idnp, zhu2022personalized, he2020lightgcn, liu2020cross}, potentially leading to inappropriate aggregation of heterogeneous neighbors~\cite{cao2022cross, cao2022disencdr}, 
we derive deterministic user- or item-specific representations that bridge two-hop neighbors in respective categories, i.e., connecting users with users.
Note that the following steps are domain-agnostic,
we omitted domain indicators $S$ and $T$ for simplicity.
Using a $k$-layer GCN, the following node representations $\tilde{\mathbf{E}}^{(k)}$ are computed:
\begin{equation}
\tilde{\mathbf{E}}^{(k)} \leftarrow \left\{
    \begin{aligned}
    &\mathbf{\Phi}^{(1)} ( \tilde{\mathbf{E}}^{(1)} \mathrel{\Vert} \boldsymbol{E}), &\mathrm{if}\, k = 1 \\
    &\mathbf{\Phi}^{(k)} ( \tilde{\mathbf{E}}^{(k-1)} \mathrel{\Vert} \tilde{\mathbf{E}}^{(k-2)}), &\mathrm{if}\, k\geq 2 \\
    \end{aligned}
    \right.
\end{equation}
\begin{equation}\label{eq: gcn}
    \text{with } \tilde{\mathbf{E}}^{(1)} := \left\{
    \begin{aligned}
        \tilde{\mathbf{U}}^{(1)} & = 
        \overline{\boldsymbol{A}} \left(\mathbf{\Phi}^{(0)}\left(\overline{\boldsymbol{A}}^\top \boldsymbol{U} \boldsymbol{W}_{u}\right)\right) \boldsymbol{W}_{u^{\prime}}, \\
        \tilde{\mathbf{V}}^{(1)} & = 
        \overline{\boldsymbol{A}}^\top \left(\mathbf{\Phi}^{(0)}\left(\overline{\boldsymbol{A}} \boldsymbol{V} \boldsymbol{W}_v \right)\right) \boldsymbol{W}_{v^{\prime}},
    \end{aligned}
    \right.
\end{equation}
where $\tilde{\mathbf{E}}^{(k)} = \{\tilde{\mathbf{U}}^{(k)}$, $\tilde{\mathbf{V}}^{(k)}$\} and $\boldsymbol{E} = \{\boldsymbol{U}, \boldsymbol{V}\}$.
$\boldsymbol{U}\in \mathbb{R}^{|\mathcal{U}| \times d}$ and $\boldsymbol{V}\in \mathbb{R}^{|\mathcal{V}| \times d}$ are random initial embeddings of users and items.
$d$ is the embedding size.
$\boldsymbol{W}_u, \boldsymbol{W}_v$ are random weight parameters.
$\mathrel{\Vert}$ is vector concatenation.
$\mathbf{\Phi}^{(k)}(\cdot)$ defines a non-linear transformation at the $k$-th layer with LeakyReLU activation function.
$\overline{\boldsymbol{A}}$ represents the normalized adjacency matrix\footnote{$\overline{\boldsymbol{A}}$ indicates the edges from $u$ to $v$, thus its transpose $\overline{\boldsymbol{A}}^\top$ is from $v$ to $u$.}.
Eq.~(\ref{eq: gcn}) ensures that each node aggregates its two-hop homogeneous neighbors.
Following~\cite{he2020lightgcn}, the outputs from all $k$ layers are then concatenated to form $\tilde{\mathbf{U}} \in \mathbb{R}^{|\mathcal{U}| \times k \cdot d}$ and $\tilde{\mathbf{V}}  \in \mathbb{R}^{|\mathcal{V}| \times k \cdot d}$ for all observed users and items, respectively.

\subsection{Stochastic Latent Preference Identifier}
\label{section: identifier}
Having obtained the deterministic representations for source and target domains, our next step is to uncover the underlying preference patterns existing across both domains, i.e., cross-domain invariant preference.
Recall that our approach avoids discrete {\it explicit} matching and instead aligns the users' preferences on a {\it implicit} basis, by assuming user behaviors follow a continuous prior preference distribution $p(\mathbf{z})$.
This inherently enables the inference of latent correlations among users.

\begin{figure}
    \centering
        \includegraphics[width=\linewidth]{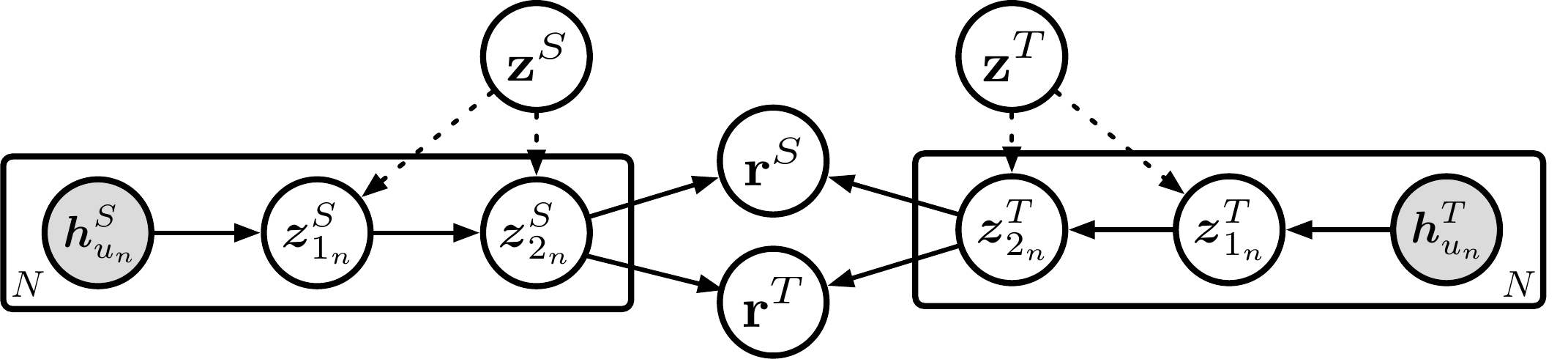}
        \caption{Graphical model for identifying the {\it Cross-domain Invariant Preference} from source domain $S$ and target domain $T$, with random $N$ users in both domains.}
        \label{fig: graphical_model}   
    \vskip -0.2in
\end{figure}

Assume a random group of user representations $\tilde{\mathbf{h}}$ are i.i.d discrete observations drawn from a generative process in the form of $p(\mathbf{h}, \mathbf{z}) = p(\mathbf{z})p(\mathbf{h}|\mathbf{z})$ over latent variables $\mathbf{z}$ for each domain.
Generally, the true posterior $p(\mathbf{z} | \mathbf{h})$ is intractable, thus needs to be approximated by a conditional posterior $q(\mathbf{z} | \tilde{\mathbf{h}})$ with the amortized inference~\cite{gershman2014amortized} (here $\tilde{\mathbf{h}} = \{ \tilde{\boldsymbol{h}}_{u_n} \}_{n=1:N} \in \tilde{\mathbf{U}}$)\footnote{The user representations $\tilde{\boldsymbol{h}}_{u_n} \in \tilde{\mathbf{U}}$ are generated from Sec. (\ref{section: encoder}). 
Here we only show source-domain user inference, omitting $S$ and $T$ for brevity. 
We also obtain the latent item representations $\tilde{\mathbf{z}}_{1, v}$ from $\tilde{\boldsymbol{h}}_{v_n} \in \tilde{\mathbf{V}}$, following a similar procedure in Sec. (\ref{subsec: infer_z1}).}.
To improve model expressivity~\cite{klushyn2019learning, vahdat2020nvae}, our implementation borrows from hierarchical priors $\mathbf{z} = \{ \mathbf{z}_{l} \}_{l=1:L}$.
That is, we represent the prior by $p(\mathbf{z}) = \prod_{l} p(\mathbf{z}_{l} | \mathbf{z}_{<l})$, and corresponding posterior by $q(\mathbf{z} | \mathbf{x}) = \prod_{l} q(\mathbf{z}_{l} | \mathbf{z}_{<l}, \tilde{\mathbf{h}})$.
In practice, we set $L=2$ to account for latent user-specific representation($l=1$) and domain-level preference($l=2$).
Upon domain-level preference $q(\mathbf{z}_{2} | \mathbf{z}_{1}, \mathbf{h})$, we can extract the cross-domain invariant preference (denoted as $\mathbf{r}$) and parameterize a predictive distribution over $N$ observations $\tilde{\mathbf{h}}$.
Thus, the source and target domains would yield a prediction $p(\mathbf{r}^{S} | \mathbf{z})$ and $p(\mathbf{r}^{T} | \mathbf{z})$, respectively.
This represents an interpretation of cross-domain invariant preference from the source/target domain perspective.
Fig.~(\ref{fig: graphical_model}) shows the graphical model.


\subsubsection{Inference of $q(\mathbf{z}_{1} | \tilde{\mathbf{h}})$}
\label{subsec: infer_z1}
Assume $p(\mathbf{z}_{1}) \sim \mathcal{N}(0, \mathbf{I})$,
we conclude a Gaussian posterior $q(\mathbf{z}_1 | \tilde{\mathbf{h}} ) = \mathcal{N}(\boldsymbol{\mu}_{1}, \text{diag}(\boldsymbol{\Sigma}_{1}))$ by computing the sufficient statistics with respect to $N$ users.
The reparameterization trick~\cite{kingma2013auto} enables us to sample the latent user representation as
\begin{equation}\label{eq: latent_node}
    \tilde{\mathbf{z}}_{1} = \boldsymbol{\mu}_{1} + \boldsymbol{\Sigma}_{1} \odot \boldsymbol{\epsilon}_{1},
    \; \text{with } \boldsymbol{\epsilon}_{1} \sim \mathcal{N}(0, \mathbf{I}), \\
\end{equation}
Specifically, $\boldsymbol{\mu}_{1}$ and $\boldsymbol{\Sigma}_{1}$ are parameterized with a multilayer perceptron~(MLP) $f_{(\cdot)}$, 
where $\boldsymbol{\mu}_{1} = f_{\boldsymbol{\mu}_{1}}( \tilde{\mathbf{h}}) \in \mathbb{R}^{N \times k \cdot d}, 
\boldsymbol{\Sigma}_{1} =f_{\boldsymbol{\Sigma}_{1}}(\tilde{\mathbf{h}})\in \mathbb{R}^{N \times k \cdot d}$.
$\odot$ denotes element-wise multiplication.


\subsubsection{Inference of $q(\mathbf{z}_2 | \mathbf{z}_1, \tilde{\mathbf{h}})$}\label{subsec: infer_z2}

Providing $q(\mathbf{z}_{1} | \tilde{\mathbf{h}})$, we further capture the domain-level user preference $p(\mathbf{z}_{2} | \mathbf{z}_{1})$ based on latent user representations $\tilde{\mathbf{z}}_{1}$.
For either domain, we approximate the posterior with $q(\mathbf{z}_2 | \mathbf{z}_{1}, \tilde{\mathbf{h}}) = \mathcal{N}(\boldsymbol{\mu}_{2}, \boldsymbol{\Sigma}_{2})$ using two MLPs on $\tilde{\mathbf{z}}_{1}$.
Akin to Eq.~(\ref{eq: latent_node}), we sample domain-level preference with the reparameterization trick:
\begin{equation}
    \begin{aligned}
        \tilde{\boldsymbol{z}}_{2} = \boldsymbol{\mu}_{2} + \boldsymbol{\Sigma}_{2} \odot \boldsymbol{\epsilon}_{2}, \; \text{with } \boldsymbol{\epsilon}_{2} \sim \mathcal{N}(0, \mathbf{I}),
    \end{aligned}
\end{equation}
where $\boldsymbol{\mu}_{2} = f_{\boldsymbol{\mu}_{2}}(\tilde{\mathbf{z}}_{1}) \in \mathbb{R}^{N \times d}, 
\boldsymbol{\Sigma}_{2} = f_{\boldsymbol{\Sigma}_{2}}(\tilde{\mathbf{z}}_{1}) \in \mathbb{R}^{N \times d}$.


\subsection{Distributional Preference Matching}
\label{subsec: match}
\subsubsection{Parameterizing $p(\mathbf{r} | \mathbf{z})$}
\label{subsec: domain pref}
With $\tilde{\boldsymbol{z}}_{2}^{S}$ and $\tilde{\boldsymbol{z}}_{2}^{T}$ estimated from both domains, we define two conditional predictive distributions, $p(\mathbf{r}^{S} | \tilde{\boldsymbol{z}}_{2}^{S}, \tilde{\boldsymbol{z}}_{2}^{T})$ and $p(\mathbf{r}^{T} | \tilde{\boldsymbol{z}}_{2}^{S}, \tilde{\boldsymbol{z}}_{2}^{T})$, to model the cross-domain invariant preference $\mathbf{r}^{S}$ and $\mathbf{r}^{T}$, described by source and target domain, respectively.
Following, we create a shared latent space to align such two interpretations, by minimizing the Jenson-Shannon~(JS) divergence between them\footnote{The symmetric JS divergence allows a bi-directional transfer. It means that the cross-domain transfer can be performed from source domain $S$ to target domain $T$, and from target domain $T$ to source domain $S$ as well.}.

Given latent domain-level preferences $\tilde{\boldsymbol{z}}_{2}^{S}$ and $\tilde{\boldsymbol{z}}_{2}^{T}$,
we distinguish source-driven interpretations and target-driven ones using the attention mechanism~\cite{vaswani2017attention} by varying the queries.
In the source perspective, we first apply a linear transformation on the concatenation of $\tilde{z}_{2}^{S}$ and $\tilde{z}_{2}^{T}$ to form a source-driven representation
    $ \dot{\mathbf{r}}_u^{S} = \boldsymbol{W}_{\dot{\mathbf{r}}}(\tilde{z}_{2}^{S} \, \Vert \, \tilde{z}_{2}^{T}) + \boldsymbol{b}_{\dot{\mathbf{r}}} \in \mathbb{R}^{N \times k \cdot d} $.
Then, the source-driven correlations within $\dot{\mathbf{r}}_u^{S}$ are encoded by a multi-head self-attention.
Similarly, reversing the concatenation would produce a target-driven $\tilde{\mathbf{r}}_u^{T}$.
Following, we parameterize $p(\mathbf{r}^{S} | \tilde{\boldsymbol{z}}_{2}^{S}, \tilde{\boldsymbol{z}}_{2}^{T})$ and $p(\mathbf{r}^{T} | \tilde{\boldsymbol{z}}_{2}^{S}, \tilde{\boldsymbol{z}}_{2}^{T})$ by assuming their Gaussian forms and computing the sufficient statistics $\{ \boldsymbol{\mu}^{S}_{\mathbf{r}}, \boldsymbol{\Sigma}^{S}_{\mathbf{r}} \}$ and $\{ \boldsymbol{\mu}^{T}_{\mathbf{r}}, \boldsymbol{\Sigma}^{T}_{\mathbf{r}} \}$, where $\boldsymbol{\mu}_{\mathbf{r}}^{S} = f_{\boldsymbol{\mu}_{\mathbf{r}}^{S}}(\tilde{\mathbf{r}}_u^{S})$, $\boldsymbol{\Sigma}_{\mathbf{r}}^{S} = f_{\boldsymbol{\Sigma}_{\mathbf{r}}^{S}}(\tilde{\mathbf{r}}_u^{S})$.
Note that $\boldsymbol{\mu}_{\mathbf{r}}^{T} $ and $\boldsymbol{\Sigma}_{\mathbf{r}}^{T}$ are computed in the same way.

\subsubsection{Alignment of $p(\mathbf{r} | \mathbf{z})$}
\label{subsec: align}
Further, aligning two predictive distributions, i.e., $p(\mathbf{r}^{S} | \tilde{\boldsymbol{z}}_{2}^{S}, \tilde{\boldsymbol{z}}_{2}^{T})$ and $p(\mathbf{r}^{T} |\tilde{\boldsymbol{z}}_{2}^{S}, \tilde{\boldsymbol{z}}_{2}^{T})$ is performed in light of minimizing the JS divergence, to ensure consistent preference in the latent space.
It is feasible as their concrete function forms have been specified.
We denote the {\it distributional preference matching} objective by $\mathcal{L}_{m}$, such that
\begin{equation}\label{eq: KL}
    \begin{aligned}
        \mathcal{L}_{m} = 
        \frac{1}{2} & \left\{ D_{KL} \left( p(\mathbf{r}^{S} | \tilde{\boldsymbol{z}}_{2}^{S}, \tilde{\boldsymbol{z}}_{2}^{T}), p(\mathbf{r}^{T} |\tilde{\boldsymbol{z}}_{2}^{S}, \tilde{\boldsymbol{z}}_{2}^{T})\right) \right. \\ 
        & \left. + D_{KL}\left( p(\mathbf{r}^{T} |\tilde{\boldsymbol{z}}_{2}^{S}, \tilde{\boldsymbol{z}}_{2}^{T}), p(\mathbf{r}^{S} | \tilde{\boldsymbol{z}}_{2}^{S}, \tilde{\boldsymbol{z}}_{2}^{T}) \right) \right\}
    \end{aligned}
\end{equation}
where $D_{KL}(\cdot, \cdot)$ is the Kullback-Leibler~(KL) divergence between two distributions.
Matching the predictive distribution from user groups eliminates the need for specify accurate {\it explicit} matching, which is beneficial for cold-start NOCDR.

\subsection{Predictive Optimization}
Having established the preference matching process, we now detail the objectives constraining latent user/item representations and domain-level preferences upon predicting the user-item interactions.
Our predictive objectives are inspired by Variational Information Bottleneck~(VIB)~\cite{alemi2016deep} that derive the latent representations as minimal sufficient statistics for predicting user-item interactions.
Specifically, we present {\it user-specific optimizers} to constrain the user- and item-specific representations, i.e. $\tilde{\mathbf{z}}_{1}$; and {\it domain-specific optimizer} to constrain the domain-level preference representations, i.e., $\tilde{\mathbf{z}}_{2}$.

\subsubsection{User-specific Optimizer}
We expect the latent user representations to (a) remove irrelevant features from user and item representations, and (2) preserve maximum information provided by user-item interactions.
Intuitively, this motivates the use of VIB in optimizing the latent user-specific $\tilde{\mathbf{z}}_{1, u}$ and item-specific representations $\tilde{\mathbf{z}}_{1, v}$.
We set up a {\it user-specific optimizer}\footnote{{\it user-specific optimizer} optimizes for latent node-level representations in Sec. (\ref{subsec: infer_z1}). Thus, item representations are actually handled as well.} for source domain $S$ and target domain $T$, respectively.
Below we illustrate $S$ only for brevity.
The VIB-informed objective is formulated as
\begin{equation}
\label{eq: infor}
\begin{aligned}
    \max 
    \left( I(\tilde{\mathbf{z}}_{1, u}^{S}; \, \boldsymbol{A}^{S}) + I(\tilde{\mathbf{z}}_{1, v}^{S}; \, \boldsymbol{A}^{S}) \right), \\
    \text{s.t.}\; \min
    \left( I(\boldsymbol{U}^S; \, \tilde{\mathbf{z}}_{1, u}^{S}) + I(\boldsymbol{V}^S; \, \tilde{\mathbf{z}}_{1, v}^{S}) \right)
\end{aligned}
\end{equation}
where $I(\cdot \,; \cdot)$ measures the mutual information.
$\boldsymbol{U}^S$, $\boldsymbol{V}^S$ are initial embeddings of users and items in the source domain. 
$\boldsymbol{A}^{S}$ is the interaction adjacency matrix in $\mathcal{G}^{S}$, approximating information of interest from user-item interactions in the source domain.
The details are covered in Eq.~(\ref{eq: min}).
Moreover, the latent user representation $\tilde{\mathbf{z}}_{1, u}^{S}$ and item representation $\tilde{\mathbf{z}}_{1, v}^{S}$ are independent in terms of different generative processes.
This enables us to further transform $I(\tilde{\mathbf{z}}_{1, u}^{S};\boldsymbol{A}^{S}) + I(\tilde{\mathbf{z}}_{1, v}^{S};\boldsymbol{A}^{S})$ into $I(\tilde{\mathbf{z}}_{1, u}^{S}, \tilde{\mathbf{z}}_{1, v}^{S};\boldsymbol{A}^{S})$ with the mutual information chain rule~\cite{cao2022cross}.
Equivalently, Eq.~(\ref{eq: infor}) can be translated to a constrained optimization objective denoted by $\mathcal{L}_{u}^{S}$, such that 
\begin{equation}
    \label{eq: opt}
    \begin{aligned}
    \mathcal{L}_{u}^{S} = - (
    \underbrace{I( \tilde{\mathbf{z}}_{1, u}^{S}, \tilde{\mathbf{z}}_{1, v}^{S}; \boldsymbol{A}^{S})}_{\text{in-domain interactions \ding{172}}} \underbrace{-\beta_{u}^{S} \cdot I(\boldsymbol{U}^S; \tilde{\mathbf{z}}_{1, u}^{S})}_{\text{users \ding{183}}} \underbrace{-\beta_{v}^{S} \cdot I(\boldsymbol{V}^S; \, \tilde{\mathbf{z}}_{1, v}^{S})}_{\text{items \ding{184}} })
    \end{aligned}
\end{equation}
where $\beta_{u}^{S}, \beta_{v}^{S}$ are Lagrangian multipliers for users and items in the source domain.
\ding{183}, \ding{184} are intractable mutual information terms~\cite{hwang2020variational} that encourage $\tilde{\mathbf{z}}_{1, u}^{S}$ and $\tilde{\mathbf{z}}_{1, v}^{S}$ to be maximally compressive about $\boldsymbol{U}^{S}$ and $\boldsymbol{V}^{S}$.
Our practice is to minimize variational approximations~\cite{farnia2016minimax} instead.
Taking $I(\boldsymbol{U}^S; \, \tilde{\mathbf{z}}_{1, u}^{S})$ as an example, its variational upper bound~\cite{poole2019variational} is given by
\begin{equation}
\label{eq: max}
    \begin{aligned}
        I(\boldsymbol{U}^S; \tilde{\mathbf{z}}_{1, u}^{S})
        & \leq D_{KL}\left( q(\tilde{\mathbf{z}}_{1, u}^{S} | \boldsymbol{U}^S), \, p(\mathbf{z}_{1, u}^{S}) \right) \\
        & = D_{KL}\left( q(\tilde{\mathbf{z}}_{1, u}^{S} | \tilde{\mathbf{h}}^S_{u}), \, p(\mathbf{z}_{1, u}^{S}) \right) \\
    \end{aligned}
\end{equation}
where both $q(\tilde{\mathbf{z}}_{1, u}^{S} | \tilde{\mathbf{h}}^S_{u})$ and $p(\mathbf{z}_{1})$ take the form of Gaussian as specified in Sec.~(\ref{subsec: infer_z1}).
Essentially,~\ding{172} assesses the ability to reconstruct observed user-item interactions.
Following~\cite{truong2021bilateral}, we can parameterize the conditional likelihood as
\begin{equation}
    \label{eq: min}
    \begin{aligned}
        I(\tilde{\mathbf{z}}_{1, u}^{S}, \tilde{\mathbf{z}}_{1, v}^{S}; \, \boldsymbol{A}^{S}) & \geq \mathbb{E}_{Q_{U}\, Q_{V}} \left[ \log p(\boldsymbol{A^S} | \tilde{\mathbf{z}}_{1, u}^{S}, \tilde{\mathbf{z}}_{1, v}^{S}) \right] \\
                        & = \sum_{u_i}^{\boldsymbol{U}^S} \sum_{v_j}^{\boldsymbol{V}^S} \log g( \tilde{\boldsymbol{z}}^{S}_{1, u_i}, \tilde{\boldsymbol{z}}^{S}_{1, v_j}) 
    \end{aligned}
\end{equation}
where $Q_U, Q_V$ abbreviate $q(\tilde{\mathbf{z}}^{S}_{1, u} | \tilde{\mathbf{h}}^S_{u})$ and $q(\tilde{\mathbf{z}}^{S}_{1, v} | \tilde{\mathbf{h}}^S_{v})$. 
$\tilde{\boldsymbol{z}}^S_{u_i}$ and $\tilde{\boldsymbol{z}}^S_{v_j}$ are latent representations of user $u_i$ and item $v_j$.
$g(\cdot, \cdot)$ can be any differentiable function~\cite{truong2021bilateral} in terms of the specific prediction task, e.g., inner product.
In this study, we predict the next interaction item and adopt binary cross entropy~(BCE) as the objective.
Since Eq.~(\ref{eq: opt}, \ref{eq: max}, \ref{eq: min}) also hold for target domain $T$,
we similarly employ $\mathcal{L}_{u}^{T}$ to optimize for target domain $T$.

\subsubsection{Domain-specific Optimizer}
NOCDR targets developing transferable user representations to be effective in another domain.
While derived from random groups of users, the {\it domain-level preference} ought to encode behavioral universality of the user population, implying stronger generalization for cross-domain predictions.
To this end, we apply another two VIB objectives for $\tilde{\mathbf{z}}_{2}^{S}$ and $\tilde{\mathbf{z}}_{2}^{T}$.
In doing so, the source representation $\tilde{\mathbf{z}}_{2}^{S}$ is optimized to be a compact form of $\boldsymbol{U}^{S}$ while also being able to predict target domain interactions $\boldsymbol{A}^{T}$ accurately; and vice versa for the target representation $\tilde{\mathbf{z}}_{2}^{T}$.

Now consider the joint optimization\footnote{As mentioned in Sec.~(\ref{subsec: align}), this can be viewed as a symmetric bi-directional transfer.} of both domains.
Denote the interaction information by $\boldsymbol{A}^{ST}$, a joint optimization has the following VIB-based objective:
\begin{equation}
\label{eq: vib_domain}
\begin{aligned}
    \max 
    \left( I(\tilde{\mathbf{z}}_{2}^{S}; \, \boldsymbol{A}^{ST}) + I(\tilde{\mathbf{z}}_{2}^{T}; \, \boldsymbol{A}^{ST}) \right), \\
    \text{s.t.}\; \min
    \left( I(\boldsymbol{U}^S; \, \tilde{\mathbf{z}}_{2}^{S}) + I(\boldsymbol{U}^{T}; \, \tilde{\mathbf{z}}_{2}^{T}) \right)
\end{aligned}
\end{equation}
where $ \boldsymbol{A}^{ST} = 
    \begin{bmatrix}
    \boldsymbol{A}^S & 0\\ 0 & \boldsymbol{A}^T
    \end{bmatrix}$ is a diagonal adjacency matrix.


Similarly, we rewrite $I(\tilde{\mathbf{z}}_{2}^{S}; \boldsymbol{A}^{ST}) + I(\tilde{\mathbf{z}}_{2}^{T}; \boldsymbol{A}^{ST})$ to $I(\tilde{\mathbf{z}}_{2}^{S}, \tilde{\mathbf{z}}_{2}^{T}; \boldsymbol{A}^{ST})$, provided that $\tilde{\mathbf{z}}_{2}^{S}$ and $\tilde{\mathbf{z}}_{2}^{T}$ are independent derivations.
We thus have an equivalent constrained optimization objective for {\it domain-level preference} as
\begin{equation}
\mathcal{L}_{d} = - (
    \underbrace{I(\tilde{\mathbf{z}}_{2}^{S}, \tilde{\mathbf{z}}_{2}^{T}; \, \boldsymbol{A}^{ST})}_{\text{cross-domain interactions\ding{175}}} \underbrace{-\beta_{2}^{S} \cdot I(\boldsymbol{U}^S; \, \tilde{\mathbf{z}}_{2}^{S})}_{\text{source users \ding{186}}} \underbrace{-\beta_{2}^{T} \cdot I(\boldsymbol{U}^{T}; \, \tilde{\mathbf{z}}_{2}^{T})}_{\text{target users \ding{187}} })
\end{equation}
where $\beta_{2}^{S}, \beta_{2}^{T}$ are Lagrangian multipliers.
Variational approximations of~\ding{186} and \ding{187} are in the same way as~\ding{183} and \ding{184}.
\begin{equation}
    \begin{aligned}
        I(\boldsymbol{U}^S ; \, \tilde{\mathbf{z}}_{2}^{S}) & + I(\boldsymbol{U}^T; \, \tilde{\mathbf{z}}_{2}^{T}) \\
        & \leq D_{KL} \left( q(\tilde{\mathbf{z}}_{2}^{S} | \boldsymbol{U}^S), \, p(\mathbf{z}_{2}^{S} | \mathbf{z}_{1, u}^{S}) \right) \\
        & \qquad \qquad + D_{KL}\left( q(\tilde{\mathbf{z}}_{2}^{T} | \boldsymbol{U}^T), \, p(\mathbf{z}_{2}^{T} | \mathbf{z}_{1, u}^{T}) \right) \\
        & = D_{KL}\left( q(\tilde{\mathbf{z}}_{2}^{S} | \tilde{\mathbf{z}}_{1, u}^{S}, \tilde{\mathbf{h}}^S_{u}), \, p(\mathbf{z}_{2}^{S} | \mathbf{z}_{1, u}^{S}) \right) \\
        & \qquad \qquad + D_{KL}\left( q(\tilde{\mathbf{z}}_{2}^{T} | \tilde{\mathbf{z}}_{1, u}^{T}, \tilde{\mathbf{h}}^T_{u}), \, p(\mathbf{z}_{2}^{T} | \mathbf{z}_{1, u}^{T}) \right) \\
    \end{aligned}
\end{equation}
As aforementioned, we expect the latent {\it domain-level} preference to function effectively in both domains, i.e., source users can be adapted to the target domain, and vice versa.
We thus empirically approximate~\ding{175} as with Eq.~(\ref{eq: min}) using the predictions of users with all the items in both domains,
\begin{equation}
    \begin{aligned}
        I(\boldsymbol{R}_{s}, \boldsymbol{R}_{t}; \boldsymbol{Z})
        \geq \sum_{v_j}^{\boldsymbol{A}^{ST}} 
        & \{ \sum_{u_i}^{\boldsymbol{U^{S}}} \log g( \tilde{\boldsymbol{z}}^{S}_{2, u_i}, \tilde{\boldsymbol{z}}_{1, v_j}) \\
        & + \sum_{u_i}^{\boldsymbol{U^{T}}} \log g( \tilde{\boldsymbol{z}}^{T}_{2, u_i}, \tilde{\boldsymbol{z}}_{1, v_j}) \}
    \end{aligned}
\end{equation}
Analogously, the differentiable function $g(\cdot, \cdot)$ is set to BCE for predicting the user-item interactions.
Summarizing all objectives, we end up with the following,
\begin{equation}
    \mathcal{L} = \mathcal{L}_{m} + \mathcal{L}_{d}+ \mathcal{L}_{u}^{S} + \mathcal{L}_{u}^{T} 
\end{equation}
where $\mathcal{L}_{m}$ describes {\it distributional preference matching} in Sec. (\ref{subsec: match}),
$\mathcal{L}_{d}$ specifies the cross-domain predictive objective in {\it domain-specific optimizer}.
$\mathcal{L}_{u}^{S}$ and $\mathcal{L}_{u}^{T}$ refers to {\it user-specific optimizer} for source and target users, respectively.

\begin{table}[]
    \caption{The Statistics of Datasets.}
    \label{tab: data}
    \vskip -.05in
    \centering
    \resizebox{\linewidth}{!}{%
    \begin{tabular}{l|c|c|c|c|c|c|c|c}
    \hline
    scenarios  & users \# & N-O users \# & items \# & interactions & validation & test   & new users \# & ave length \\ 
    \hline
    Cellphone  & 27,519   & 11,182       & 9,481    & 156,581      & 6,322      & 6,417  & 2,049   & 14.02      \\ 
    Electronic & 107,984  & 100,127      & 40,460   & 795,799      & 15,053     & 15,199 & 2,042   & 7.95       \\ \hline \hline
    Cloth      & 41,829   & 33,972       & 17,943   & 180,545      & 3,085      & 3,156  & 990     & 5.31       \\ 
    Sport      & 27,328   & 19,471       & 12,656   & 160,482      & 3,546      & 3,589  & 981     & 8.24       \\ \hline \hline
    Game       & 25,025   & 23,288       & 12,319   & 144,305      & 1,304      & 1,381  & 226     & 6.20       \\ 
    Video      & 19,457   & 17,720       & 8,751    & 145,798      & 1,458      & 1,435  & 217     & 8.23       \\ \hline  \hline
    Music      & 50,841   & 35,760       & 43,858   & 685,138      & 19,670     & 19,837 & 1,893   & 19.16      \\ 
    Movie      & 87,875   & 72,794       & 38,643   & 1,119,358    & 28,876     & 28,589 & 1,885   & 15.38      \\ 
    \hline
    \end{tabular}
    }
\end{table}

\subsubsection{Time Complexity}
The time complexity for one domain is given by:
1) The Deterministic Graph Encoder (Sec. (\ref{section: encoder})):
$\mathcal{O}(k \times (|\mathcal{U}|+|\mathcal{V}|))$, $|\mathcal{U}|$ and $|\mathcal{V}|$ are the number of user/item nodes in one domain, $k$ is the number of layers;
2) The item-specific inference (Sec. (\ref{subsec: infer_z1})): $\mathcal{O}(|\mathcal{V}|)$;
3) The user-specific inference (Sec. (\ref{subsec: infer_z1})):
$\mathcal{O}(N)$, where $N$ is the sampling size of the random user group;
4) The domain-level preference inference (Sec. (\ref{subsec: infer_z2})): $\mathcal{O}(N)$;
5) The multi-head attention (Sec. (\ref{subsec: align})): $\mathcal{O}((k \times d)^2 \times h)$, $h$ is the number of heads;
6) The distributional preference matching (Sec. (\ref{subsec: match}): $\mathcal{O}(N)$.

Denote the node number of the source domain by $|S|$, and the target domain by $|T|$. 
Summing all computations up gives us the overall asymptotic upper bound as $\mathcal{O}(k \cdot (|S|^2 + |T|^2) + 6N +|V^S|+|V^T| +2h \times k^2d^2)$ for two domains.
The complexity of {\it Stochastic Latent Preference Identifier} and {\it Distributional Preference Matching} scale linearly with the number of samples, whose practical computations can take advantage of the parallelization enabled by a GPU.

\begin{table*}[]
\caption{Overall performance(\%).
Improves(\%) shows the improvement of DPMCDR over the runner-up results.
* indicates a statistically significant improvement (p$<0.05$) when comparing DPMCDR and the best baseline using a paired t-test.
}
\label{tab: cell-electronic}
\centering
\resizebox{\linewidth}{!}{%
\begin{tabular}{l|ccccccc|ccccccc}
\hline
\multirow{2}{*}{Methods} & \multicolumn{7}{c|}{Cellphone}                                                                                                                                                              & \multicolumn{7}{c}{Electronic}                           \\   \cline{2-15}                                                       
                         & MRR                      & NDCG@10                  & NDCG@20                  & NDCG@30                  & HR@10                    & HR@20                    & HR@30                    & MRR                      & NDCG@10                  & NDCG@20                  & NDCG@30                  & HR@10                    & HR@20                    & HR@30                    \\
                         \hline
MF                       & 1.71($\pm$0.20)               & 1.30($\pm$0.07)               & 1.73($\pm$0.11)               & 2.18($\pm$0.21)               & 2.86($\pm$0.45)               & 5.47($\pm$1.35)               & 7.55($\pm$0.80)               & 1.67($\pm$0.42)               & 1.31($\pm$0.22)               & 1.49($\pm$0.22)               & 1.71($\pm$0.07)               & 2.60($\pm$0.90)               & 3.39($\pm$0.09)               & 4.43($\pm$0.18)               \\
Caser                    & 2.87($\pm$0.15)               & 2.81($\pm$0.34)               & 3.50($\pm$0.14)               & 3.75($\pm$0.15)               & 2.94($\pm$0.09)               & 6.30($\pm$0.18)               & 8.67($\pm$0.12)               & 1.32($\pm$0.26)               & 1.67($\pm$0.21)               & 1.97($\pm$0.20)               & 2.29($\pm$0.29)               & 1.30($\pm$0.21)               & 2.45($\pm$0.32)               & 3.14($\pm$0.08)               \\
IDNP                     & 3.94($\pm$0.14)               & 1.78($\pm$0.23)               & 2.55($\pm$0.07)               & 3.37($\pm$0.17)               & 3.63($\pm$0.12)               & 7.19($\pm$0.14)               & 10.87($\pm$0.32)              & 2.24($\pm$0.16)               & 1.24($\pm$0.22)               & 1.61($\pm$0.09)               & 2.34($\pm$0.21)               & 2.49($\pm$0.15)               & 4.71($\pm$0.08)               & 7.23($\pm$0.10)               \\
\hline
CMF                     & 3.01($\pm$0.07)               & 2.86($\pm$0.28)               & 3.36($\pm$0.39)               & 3.59($\pm$0.30)               & 6.51($\pm$2.25)               & 8.07($\pm$2.26)               & 9.64($\pm$1.26)               & 1.70($\pm$0.15)               & 1.13($\pm$0.30)               & 1.40($\pm$0.54)               & 1.57($\pm$0.25)               & 2.34($\pm$0.35)               & 2.86($\pm$0.18)               & 5.21($\pm$0.18)               \\
EMCDR-MF                 & 1.18($\pm$0.08)               & 1.43($\pm$0.22)               & 1.41($\pm$0.11)               & 1.75($\pm$0.17)               & 1.05($\pm$0.45)               & 4.95($\pm$0.90)               & 6.51($\pm$0.45)               & 1.81($\pm$0.20)               & 1.29($\pm$0.09)               & 1.36($\pm$0.21)               & 1.69($\pm$0.49)               & 0.78($\pm$0.01)               & 1.04($\pm$0.04)               & 2.60($\pm$0.18)               \\
EMCDR-NGCF               & 2.69($\pm$0.36)               & 2.92($\pm$0.27)               & 3.51($\pm$0.32)               & 4.14($\pm$0.70)               & 5.43($\pm$0.78)               & 8.81($\pm$0.49)               & 13.20($\pm$1.29)              & 3.14($\pm$0.46)               & 3.20($\pm$0.31)               & 3.93($\pm$0.55)               & 5.05($\pm$0.16)               & 5.09($\pm$0.39)               & 9.63($\pm$0.57)               & 12.11($\pm$0.47)              \\
PTUPCDR                  & 4.27($\pm$0.26)               & 4.58($\pm$0.29)               & 5.29($\pm$0.38)               & 6.73($\pm$0.68)               & 8.04($\pm$0.45)               & 12.84($\pm$0.12)              & 15.99($\pm$0.90)              & 5.94($\pm$0.09)               & 6.47($\pm$0.23)               & 7.86($\pm$0.21)               & 9.30($\pm$0.12)               & 11.30($\pm$0.09)              & 12.86($\pm$0.90)              & 14.95($\pm$0.45)              \\
DisenCDR                 & 4.34($\pm$1.26)               & 4.43($\pm$1.42)               & 5.85($\pm$0.78)               & 6.51($\pm$0.70)               & 8.76($\pm$1.05)               & 14.36($\pm$1.20)              & 17.48($\pm$1.99)              & 6.88($\pm$0.56)               & 7.06($\pm$0.67)               & 8.07($\pm$0.74)               & 8.72($\pm$0.88)               & 11.12($\pm$1.12)              & 15.09($\pm$1.45)              & 18.15($\pm$2.14)              \\
CRDIB                    & {\ul 6.33($\pm$0.34)}         & {\ul 7.32($\pm$0.33)}         & {\ul 9.26($\pm$0.44)}         & {\ul 10.49($\pm$0.43)}        & {\ul 15.03($\pm$0.37)}        & {\ul 22.72($\pm$0.83)}        & {\ul 27.50($\pm$0.86)}          & {\ul 8.48($\pm$0.21)}         & {\ul 9.65($\pm$0.25)}         & {\ul 10.76($\pm$0.29)}        & {\ul 13.05($\pm$0.28)}        & {\ul 17.19($\pm$0.44)}        & {\ul 25.53($\pm$0.57)}        & {\ul 31.61($\pm$0.53)}        \\
\hline
\textbf{DPMCDR*}         & \textbf{8.54($\pm$0.35)}      & \textbf{9.68($\pm$0.30)}      & \textbf{11.65($\pm$0.32)}     & \textbf{12.90($\pm$0.28) }    & \textbf{17.91($\pm$0.11)}     & \textbf{25.72($\pm$0.37)}     & \textbf{31.65($\pm$0.65)}     & \textbf{10.31($\pm$0.11)}     & \textbf{11.51($\pm$0.14)}     & \textbf{13.68($\pm$0.11)}     & \textbf{15.01($\pm$0.13) }    & \textbf{20.20($\pm$0.37)}     & \textbf{28.82($\pm$0.15)}         & \textbf{35.08($\pm$0.38)}              \\
Improves(\%)             & 34.91\% & 32.24\% & 25.81\% & 22.97\% & 19.16\% & 13.20\% & 15.09\% & 21.58\% & 19.27\% & 28.81\% & 15.02\% & 17.51\% & 12.89\% & 10.98\%       \\
\hline
\end{tabular}
}

\resizebox{\linewidth}{!}{%
\begin{tabular}{l|ccccccc|ccccccc}
\hline
\multirow{2}{*}{Methods} & \multicolumn{7}{c|}{Game}                                                                                 & \multicolumn{7}{c}{Video}                                                                   \\ \cline{2-15} 
                         & MRR        & NDCG@10    & NDCG@20    & NDCG@30    & HR@10      & HR@20       & \multicolumn{1}{c|}{HR@30} & MRR        & NDCG@10    & NDCG@20    & NDCG@30    & HR@10       & HR@20       & HR@30       \\ \hline
MF                       & 1.89($\pm$0.28) & 1.60($\pm$0.15) & 1.85($\pm$0.08) & 2.19($\pm$0.18) & 3.13($\pm$0.27) & 4.17($\pm$0.51)  & 6.25($\pm$0.71)                 & 2.53($\pm$0.23) & 1.98($\pm$0.17) & 2.43($\pm$0.48) & 2.86($\pm$0.42) & 2.13($\pm$0.06)  & 3.47($\pm$0.35)  & 4.81($\pm$0.07)  \\
Caser                    & 1.69($\pm$0.15) & 1.62($\pm$0.19) & 2.31($\pm$0.58) & 2.43($\pm$0.36) & 1.45($\pm$0.37) & 2.31($\pm$0.43)  & 3.55($\pm$0.43)                 & 2.05($\pm$0.03) & 1.62($\pm$0.28) & 2.44($\pm$0.38) & 2.72($\pm$0.31) & 2.16($\pm$0.34)  & 2.84($\pm$0.22)  & 3.68($\pm$0.28)  \\
IDNP                     & 1.63($\pm$0.28) & 1.90($\pm$0.28) & 2.59($\pm$0.86) & 4.96($\pm$0.74) & 5.19($\pm$1.46) & 6.12($\pm$0.26)  & 8.47($\pm$1.24)                 & 2.46($\pm$0.12) & 1.26($\pm$0.03) & 1.75($\pm$0.12) & 2.49($\pm$0.21) & 3.57($\pm$0.15)  & 4.23($\pm$0.14)  & 8.11($\pm$0.40)  \\
\hline
CMF                     & 2.97($\pm$0.36) & 2.65($\pm$0.27) & 3.63($\pm$0.36) & 4.62($\pm$0.38) & 4.47($\pm$1.07) & 6.64($\pm$1.30)  & 8.06($\pm$0.35)                 & 3.28($\pm$0.14) & 2.82($\pm$0.14) & 4.28($\pm$0.45) & 4.78($\pm$0.74) & 3.21($\pm$0.09)  & 5.68($\pm$0.80)  & 6.28($\pm$0.71)  \\
EMCDR-MF                 & 1.14($\pm$0.16) & 1.08($\pm$0.04) & 1.36($\pm$0.29) & 1.64($\pm$0.07) & 1.56($\pm$0.06) & 4.69($\pm$0.82)  & 5.99($\pm$0.85)                 & 1.13($\pm$0.28) & 1.98($\pm$0.71) & 2.16($\pm$0.71) & 2.64($\pm$0.43) & 3.57($\pm$0.59)  & 4.43($\pm$0.90)  & 6.77($\pm$0.45)  \\
EMCDR-NGCF               & 1.61($\pm$0.13) & 1.25($\pm$0.16) & 1.81($\pm$0.25) & 2.13($\pm$0.61) & 2.86($\pm$0.90) & 5.21($\pm$1.19)  & 6.77($\pm$0.80)                 & 1.90($\pm$0.21) & 1.62($\pm$0.30) & 2.34($\pm$0.16) & 2.64($\pm$0.66) & 3.65($\pm$0.19)  & 7.03($\pm$0.73)  & 8.59($\pm$1.35)  \\
PTUPCDR                  & 2.67($\pm$0.31) & 2.15($\pm$0.41) & 2.53($\pm$0.43) & 3.53($\pm$0.45) & 4.34($\pm$0.78) & 6.91($\pm$0.35)  & 8.59($\pm$0.78)                 & 2.32($\pm$0.34) & 2.07($\pm$0.08) & 2.51($\pm$0.09) & 3.95($\pm$0.16) & 4.86($\pm$0.26)  & 6.13($\pm$0.47)  & 8.69($\pm$0.77)  \\
DisenCDR                 & 2.05($\pm$0.48) & 2.13($\pm$0.36) & 2.84($\pm$0.30) & 3.14($\pm$0.38) & 4.91($\pm$0.77) & 7.69($\pm$0.39)  & 9.10($\pm$0.46)                 & 2.15($\pm$0.36) & 1.91($\pm$0.30) & 2.60($\pm$0.49) & 3.20($\pm$0.53) & 5.12($\pm$0.43)  & 7.95($\pm$0.58)  & 9.54($\pm$0.91)  \\
CRDIB                    & {\ul 4.05($\pm$0.57)}        & {\ul 4.12($\pm$0.43)}            & {\ul 5.58($\pm$0.58)}            & {\ul 6.66($\pm$0.66)}            & {\ul 7.54($\pm$0.21)}          & {\ul 13.35($\pm$0.82)}         & {\ul 18.42($\pm$1.09)}          & {\ul 4.20($\pm$0.15)}        & {\ul 4.37($\pm$0.31)}            & {\ul 5.73($\pm$0.24)}            & {\ul 6.65($\pm$0.32)}            & {\ul 7.57($\pm$0.79)}          & {\ul 12.98($\pm$0.58)}         & {\ul 17.30($\pm$1.03)}         \\
\hline
\textbf{DPMCDR*}          & \textbf{4.75($\pm$0.41)}     & \textbf{4.75($\pm$0.14)}         & \textbf{6.40($\pm$0.13)}         & \textbf{7.61($\pm$0.29)}         & \textbf{9.46($\pm$0.41)}       & \textbf{16.12($\pm$0.36)}      & \textbf{21.82($\pm$0.79)}       & \textbf{5.11($\pm$0.40)}     & \textbf{5.56($\pm$0.36)}         & \textbf{6.93($\pm$0.52)}         & \textbf{7.93($\pm$0.51)}         & \textbf{10.72($\pm$0.55)}      & \textbf{16.22($\pm$1.19)}      & \textbf{21.07($\pm$0.96)}      \\
Improves(\%)              & 17.28\% & 15.29\% & 14.70\% & 14.26\% & 25.46\% & 20.75\% & 18.46\% & 21.67\% & 27.23\% & 20.94\% & 19.25\% & 41.61\% & 24.96\% & 21.79\%        \\
\hline
\end{tabular}
}

\resizebox{\linewidth}{!}{%
\begin{tabular}{l|ccccccc|ccccccc}
\hline
\multirow{2}{*}{Methods} & \multicolumn{7}{c|}{Movie}                                                                                  & \multicolumn{7}{c}{Music}                                                                     \\ \cline{2-15} 
                         & MRR        & NDCG@10    & NDCG@20    & NDCG@30     & HR@10       & HR@20       & \multicolumn{1}{c|}{HR@30} & MRR        & NDCG@10    & NDCG@20     & NDCG@30     & HR@10       & HR@20       & HR@30       \\ \hline
MF                       & 1.28($\pm$0.24) & 0.69($\pm$0.17) & 0.95($\pm$0.39) & 1.11($\pm$0.39)  & 1.04($\pm$0.45)  & 2.08($\pm$0.45)  & 3.13($\pm$0.78)                 & 1.52($\pm$0.17) & 1.13($\pm$0.30) & 1.20($\pm$0.19)  & 1.34($\pm$0.45)  & 2.08($\pm$0.45)  & 2.86($\pm$0.45)  & 4.17($\pm$0.63)  \\
Caser                    & 1.53($\pm$0.22) & 1.03($\pm$0.25) & 1.26($\pm$0.26) & 1.51($\pm$0.34)  & 1.01($\pm$0.21)  & 3.73($\pm$0.15)  & 5.51($\pm$0.17)                 & 3.45($\pm$0.46) & 1.63($\pm$0.32) & 2.49($\pm$0.20)  & 3.08($\pm$0.28)  & 3.37($\pm$0.33)  & 6.52($\pm$0.32)  & 9.85($\pm$0.59)  \\
IDNP                     & 3.48($\pm$0.35) & 1.58($\pm$0.35) & 2.36($\pm$0.42) & 2.94($\pm$0.36)  & 3.23($\pm$0.43)  & 6.21($\pm$0.48)  & 9.01($\pm$0.30)                 & 2.45($\pm$0.13) & 2.13($\pm$0.27) & 3.48($\pm$0.22)  & 4.53($\pm$0.52)  & 5.54($\pm$0.10)  & 7.97($\pm$1.45)  & 8.14($\pm$0.42)  \\
\hline
CMF                     & 2.13($\pm$0.10) & 1.75($\pm$0.12) & 2.14($\pm$0.09) & 2.30($\pm$0.09)  & 3.04($\pm$0.26)  & 5.21($\pm$0.90)  & 5.99($\pm$0.45)                 & 2.32($\pm$0.24) & 1.69($\pm$0.39) & 3.35($\pm$0.95)  & 4.51($\pm$0.25)  & 4.69($\pm$0.35)  & 5.46($\pm$0.61)  & 6.93($\pm$0.96)  \\
EMCDR-MF                 & 2.62($\pm$0.16) & 1.01($\pm$0.05) & 2.83($\pm$0.16) & 3.77($\pm$0.16)  & 4.58($\pm$0.60)  & 8.07($\pm$0.22)  & 12.17($\pm$3.20)                & {\ul 6.75($\pm$0.21)} & 3.18($\pm$0.69) & 3.98($\pm$0.49)  & 5.63($\pm$0.62)  & 5.88($\pm$0.70)  & 6.90($\pm$0.40)  & 7.68($\pm$0.78)  \\
EMCDR-NGCF               & 2.18($\pm$0.28) & 2.62($\pm$0.14) & 3.05($\pm$0.13) & 4.55($\pm$0.20)  & 5.77($\pm$0.21)  & 9.91($\pm$0.43)  & 13.73($\pm$1.19)                & 3.28($\pm$1.41) & 3.08($\pm$0.89) & 4.80($\pm$0.50)  & 5.58($\pm$0.43)  & 6.51($\pm$0.58)  & 8.51($\pm$0.80)  & 9.19($\pm$0.45)  \\
PTUPCDR                  & 4.31($\pm$0.22) & 4.27($\pm$0.54) & 4.54($\pm$0.33) & 6.93($\pm$0.26)  & 7.47($\pm$0.33)  & 10.64($\pm$0.48) & 13.44($\pm$0.36)                & 2.58($\pm$0.20) & 2.39($\pm$0.14) & 2.39($\pm$0.29)  & 4.92($\pm$0.30)  & 4.75($\pm$0.53)  & 8.29($\pm$0.17)  & 11.58($\pm$0.33) \\
DisenCDR                 & 4.27($\pm$0.62) & 4.11($\pm$0.41) & 4.91($\pm$0.43) & 6.01($\pm$0.74)  & 7.33($\pm$0.30)  & 11.85($\pm$0.67) & 14.99($\pm$0.96)                & 3.47($\pm$0.31) & 3.06($\pm$0.26) & 3.68($\pm$0.63)  & 4.42($\pm$0.43)  & 5.72($\pm$0.31)  & 9.23($\pm$0.43)  & 12.07($\pm$0.71) \\
CRDIB                    & {\ul 6.41($\pm$0.21)}        & {\ul 6.91($\pm$0.25)}            & {\ul 8.88($\pm$0.30)}            & {\ul 10.17($\pm$0.35)}           & {\ul 13.31($\pm$0.57)}         & {\ul 21.15($\pm$0.79)}         & {\ul 27.22($\pm$1.07)}          &  6.65($\pm$0.38)        & {\ul 7.25($\pm$0.40)}            & {\ul 9.08($\pm$0.51)}            & {\ul 10.21($\pm$0.54)}           & {\ul 13.64($\pm$0.64)}         & {\ul 20.92($\pm$1.10)}         & {\ul 26.25($\pm$1.25)}         \\
\hline
\textbf{DPMCDR*}          & \textbf{8.07($\pm$0.32)}     & \textbf{8.95($\pm$0.37)}         & \textbf{11.2($\pm$0.43)}         & \textbf{12.60($\pm$0.49)}        & \textbf{16.94($\pm$0.62)}      & \textbf{25.88($\pm$0.77)}      & \textbf{32.48($\pm$1.17)}       & \textbf{8.63($\pm$0.33)}     & \textbf{9.65($\pm$0.41)}         & \textbf{11.71($\pm$0.45)}        & \textbf{12.97($\pm$0.48)}        & \textbf{17.80($\pm$0.77)}      & \textbf{25.99($\pm$0.97)}      & \textbf{31.90($\pm$1.23)}      \\
Improves(\%)              & 25.90\% & 29.52\% & 26.13\% & 24.39\% & 27.27\% & 22.36\% & 19.32\% & 28.27\% & 34.62\% & 28.96\% & 27.03\% & 30.50\% & 24.24\% & 21.52\%       \\
\hline
\end{tabular}
}

\resizebox{\linewidth}{!}{%
\begin{tabular}{l|ccccccc|ccccccc}
\hline
\multirow{2}{*}{Methods} & \multicolumn{7}{c|}{Cloth}                                                                                 & \multicolumn{7}{c}{Sport}                                                                  \\ \cline{2-15} 
                         & MRR        & NDCG@10    & NDCG@20    & NDCG@30    & HR@10       & HR@20       & \multicolumn{1}{c|}{HR@30} & MRR        & NDCG@10    & NDCG@20    & NDCG@30    & HR@10      & HR@20       & HR@30       \\ \hline
MF                       & 1.57($\pm$0.38) & 1.95($\pm$0.58) & 2.37($\pm$0.48) & 3.02($\pm$0.67) & 3.52($\pm$0.71)  & 5.93($\pm$0.89)  & 7.42($\pm$1.40)                 & 1.49($\pm$0.29) & 1.28($\pm$0.44) & 1.71($\pm$0.42) & 2.04($\pm$0.23) & 3.13($\pm$0.78) & 5.47($\pm$0.56)  & 7.29($\pm$0.45)  \\
Caser                    & 1.77($\pm$0.29) & 3.45($\pm$0.39) & 5.50($\pm$0.19) & 7.25($\pm$0.39) & 4.66($\pm$0.28)  & 6.59($\pm$0.34)  & 8.36($\pm$0.40)                 & 1.64($\pm$0.18) & 2.59($\pm$0.23) & 3.84($\pm$0.32) & 5.18($\pm$0.43) & 1.87($\pm$0.23) & 3.20($\pm$0.35)  & 4.61($\pm$0.31)  \\
IDNP                     & 2.13($\pm$0.25) & 1.92($\pm$0.32) & 2.46($\pm$0.18) & 2.89($\pm$0.14) & 4.17($\pm$0.90)  & 6.77($\pm$0.45)  & 8.07($\pm$1.19)                 & 1.66($\pm$0.18) & 1.92($\pm$0.27) & 2.48($\pm$0.27) & 3.82($\pm$0.57) & 2.34($\pm$0.36) & 3.91($\pm$0.18)  & 6.07($\pm$0.32)  \\
\hline
CMF                     & 2.46($\pm$0.31) & 1.99($\pm$0.41) & 2.73($\pm$0.34) & 3.43($\pm$0.32) & 4.07($\pm$0.19)  & 5.07($\pm$0.80)  & 7.20($\pm$0.97)                 & 1.46($\pm$0.27) & 1.37($\pm$0.44) & 1.64($\pm$0.16) & 1.97($\pm$0.25) & 3.34($\pm$0.23) & 4.01($\pm$0.08)  & 5.17($\pm$0.18)  \\
EMCDR-MF                 & 1.33($\pm$0.24) & 1.07($\pm$0.26) & 1.52($\pm$0.06) & 2.06($\pm$0.21) & 2.86($\pm$0.63)  & 4.69($\pm$0.56)  & 7.29($\pm$0.45)                 & 2.66($\pm$0.39) & 2.06($\pm$0.30) & 3.38($\pm$1.13) & 4.06($\pm$0.19) & 4.17($\pm$0.19) & 7.16($\pm$1.07)  & 10.54($\pm$1.23) \\
EMCDR-NGCF               & 1.96($\pm$0.79) & 1.77($\pm$0.41) & 2.12($\pm$0.39) & 2.69($\pm$0.23) & 3.91($\pm$0.56)  & 5.47($\pm$0.45)  & 8.59($\pm$0.82)                 & 2.00($\pm$0.79) & 1.61($\pm$0.38) & 2.37($\pm$0.19) & 2.91($\pm$0.13) & 3.65($\pm$0.36) & 7.55($\pm$0.63)  & 9.38($\pm$0.27)  \\
PTUPCDR                  & 2.96($\pm$0.05) & 2.53($\pm$0.30) & 3.52($\pm$0.32) & 4.27($\pm$0.27) & 5.90($\pm$0.40)  & 8.86($\pm$0.90)  & 11.21($\pm$0.19)                & 3.63($\pm$0.19) & 3.06($\pm$0.08) & 3.02($\pm$0.16) & 4.08($\pm$0.02) & 5.63($\pm$0.13) & 8.80($\pm$0.36)  & 11.45($\pm$0.74) \\
DisenCDR                 & 2.99($\pm$0.47) & 2.83($\pm$0.55) & 3.84($\pm$0.37) & 4.53($\pm$0.32) & 6.02($\pm$0.56)  & 10.09($\pm$0.24) & 13.29($\pm$0.64)                & {\ul 4.07($\pm$0.24)} & 3.91($\pm$0.20) & 4.80($\pm$0.15) & 5.45($\pm$0.28) & 6.99($\pm$0.18) & 10.53($\pm$1.09) & 13.57($\pm$1.63) \\
CRDIB              & {\ul 4.74($\pm$0.77)}        & {\ul 5.24($\pm$0.86)}            & {\ul 6.66($\pm$1.09)}            & {\ul 7.55($\pm$1.13)}            & {\ul 10.56($\pm$1.41)}         & {\ul 16.21($\pm$1.32)}         & {\ul 20.40($\pm$1.49)}          &  4.01($\pm$0.21)        & {\ul 4.33($\pm$0.20)}            & {\ul 5.60($\pm$0.21)}            & {\ul 6.37($\pm$0.25)}            & {\ul 8.76($\pm$0.16)}          & {\ul 13.80($\pm$0.36)}         & {\ul 17.41($\pm$0.46)}         \\
\hline
\textbf{DPMCDR*}          & \textbf{5.62($\pm$0.41)}     & \textbf{6.18($\pm$0.41)}         & \textbf{7.60($\pm$0.35)}         & \textbf{8.43($\pm$0.36)}         & \textbf{11.60($\pm$0.43)}      & \textbf{17.32($\pm$0.59)}      & \textbf{21.15($\pm$0.21)}       & \textbf{4.34($\pm$0.22)}     & \textbf{4.68($\pm$0.33)}         & \textbf{5.98($\pm$0.25)}         & \textbf{6.98($\pm$0.53)}         & \textbf{9.76($\pm$0.34)}       & \textbf{14.95($\pm$0.38)}      & \textbf{19.18($\pm$0.77)}      \\
Improves(\%)              & 18.57\% & 17.94\% & 14.11\% & 11.66\% & 9.85\% & 6.85\% & 3.68\% & 6.73\% & 8.08\% & 6.79\% & 9.58\% & 11.42\% & 8.33\% & 10.17\%       \\
\hline
\end{tabular}
}
\end{table*}

\section{Experiments}
We will examine the following research questions:
\begin{itemize}
    \item[RQ1)]
    Can DPMCDR perform NOCDR for cold-start users and outperform other state-of-the-art methods?
    \item[RQ2)] 
    Can DPMCDR also demonstrate competitive performance with state-of-the-art methods in OCDR settings?
    \item[RQ3)] 
    How does each module of DPMCDR contribute to model performance?
    \item[RQ4)] 
    How do model parameters affect experimental results?
\end{itemize}
\subsection{Experimental Setup}
\subsubsection{Datasets}
We conduct extensive experiments on a large-scale real-world dataset: Amazon Review Data\footnote{https://nijianmo.github.io/amazon/index.html}.
We evaluate DPMCDR and comparison methods for eight categories in this dataset.
In our experiments, we pair them up and construct four CDR scenarios: Cellphone-Electronic, Cloth-Sport, Game-Video, and Music-Movie.
Tab.~(\ref{tab: data}) details the information of each scenario.

\subsubsection{Methods for comparison}
We include eight baselines for empirical comparisons, including both single-domain recommendation~(SDR) and cross-domain recommendation~(CDR) models.
For SDR models, we include three methods: \textbf{MF}, \textbf{Caser}\cite{tang2018personalized}, and \textbf{IDNP}\cite{du2023idnp}.
Neither of these methods involves cross-domain adaptation, so we evaluate them by training in one domain and conducting cold-start prediction in another domain.
We further introduce five state-of-the-art models proposed to tackle the CDR task: \textbf{CMF}~\cite{singh2008relational}, \textbf{EMCDR}~\cite{man2017cross} (with two variants implemented with different encoders: \textbf{EMCDR-MF} and \textbf{EMCDR-NGCF}), \textbf{DisenCDR}~\cite{cao2022disencdr}, \textbf{PTUPCDR}~\cite{zhu2022personalized}, and \textbf{CDRIB}~\cite{cao2022cross}.
The reported results of CDR models are obtained upon performing cross-domain adaptation during training.
DisenCDR and CRDIB support bi-directional transfer as with DPMCDR.
They obtain the results for both domains within a single training process, whilst other CDR methods need to be trained twice for different source domains.


\subsubsection{Evaluation Setting}
Following common practices~\cite{cao2022disencdr, cao2022cross}, items with fewer than 10 interactions and users with less than 5 interactions are filtered out.
Recall that we focus on NOCDR predictions for cold-start users, we use non-overlapping users for training.
Following the full ranking principle~\cite{krichene2020sampled}, 20\% of overlapping users are randomly sampled to form the validation and testing sets, aligning with conventional settings in the baselines~\cite{cao2022cross, cao2022disencdr}.
Our evaluation simulates a cold-start scenario.
The ground-truth user-item interactions of evaluated users are available for measuring performance only.

\subsubsection{Metrics}
The prediction concerns the items that will be interacted with, we compare all the models in terms of three commonly-used metrics: MRR~(Mean Recall Ratio), HR~(Hit Ratio)@K with K$=\left\{10, 20, 30 \right\}$, and Normalized Discounted Cumulative Gain~(NDCG)@K with K varying in $\left\{10, 20, 30 \right\}$.

\subsubsection{Implementation}
All models are trained on an internal server equipped with two Intel Xeon E5-2697 v2 CPUs, a single NVIDIA RTX A5000 GPU and 768 GB of memory. 
Each method in comparison is implemented based on the official open-source code and is adapted for our evaluation setting.
We use the best (hyper-)parameters of all baseline models as the official implementation/description.
We report the results of all models in runs with five random seeds to minimize the impact of random noise.
To implement DPMCDR, we fix $k$, the layers of {\it Deterministic Graph Encoder} to 3.
For $f_{\boldsymbol{\mu}_{1}}$, we utilize a 1-layer MLP with Leaky-ReLU activation function, while for $f_{\boldsymbol{\Sigma}_{1}}$, we employ a 1-layer MLP with Softmax activation function.
Furthermore, $f_{\boldsymbol{\mu}_{2}}$, $f_{\boldsymbol{\Sigma}_{2}}$, $f_{\boldsymbol{\mu}_{\mathbf{r}}^{S}}$, $f_{\boldsymbol{\Sigma}_{\mathbf{r}}^{S}}$, $f_{\boldsymbol{\mu}_{\mathbf{r}}^{T}}$, and $f_{\boldsymbol{\Sigma}_{\mathbf{r}}^{T}}$ are implemented using 1-layer MLPs with ReLU activation function.
The multi-head attention is with 2 heads.
The dropout rate is searched within \{0, 0.1, 0.2, 0.3, 0.4, 0.5\} and is set to 0.3 for the best performance.
We search for the embedding size $d$ within \{16, 32, 64, 128\}.
We choose the best warmup epoch from \{10, 20, 30, 40, 50\}.
For simplicity, all Lagrangian multipliers $\beta$ are set to the same value, and they are searched for from \{0.5, 1, 1.5, 2, 2.5, 3\}.
We search the number of randomly selected users $N$ in {\it Stochastic Latent Preference Identifier} from \{64, 128, 256, 512, 1024\}.
The batch size is fixed to 1024.
We use Adam with a learning rate of 1e-3 and weight decay of 1e-6.
DPMCDR converges on all four tasks after at most 70 training epochs.

\subsection{Performance Analysis (RQ 1)}
Table~(\ref{tab: cell-electronic}) summarizes the overall performance for four CDR datasets.
All metrics with the best results are \textbf{bolded}; those with the second-best results are \underline{underlined}.
DPMCDR consistently outperforms all the methods in comparison with all evaluation metrics on four CDR tasks.
Moreover, the model performances support our design choices, not only for DPMCDR but also across a range of baselines.
Our discussion is framed by the following perspectives.

\subsubsection*{Powerful Graph Encoder matters}
As for SDR methods, 
MF is unable to capture user preferences accurately in all scenarios.
Caser and IDNP organize consecutive user-item interaction records as sequences, and leverage sequential patterns to capture user preferences therefrom.
For CDR methods, EMCDR-NGCF exhibits substantial improvements over EMCDR-MF despite following the same adaptation procedure, because they employ different embedding backbones.
It is widely recognized that NGCF is more expressive than MF since NGCF encodes higher-order connectivity from user-item interaction graphs using powerful graph neural networks.
This also motivates the use of GCN in DPMCDR, which encodes deterministic user/item representations from interaction data.

\subsubsection*{Latent User Correlations benefit Cross-domain Prediction}
Ideally, it is beneficial to exploit and identify the preference commonality shared by user behaviors.
The probabilistic modeling may provide a pathway, particularly with a limited size of interactions.
Observe that Caser requires extensive interaction sequences to work effectively, like with Music and Movie review data.
IDNP, on the other hand, assumes latent correlations between sequences and derives the intrinsic user preferences from a functional perspective.
In the case of limited interactions~(e.g., Game-Video), CMF connects the general user embedding between domains to reduce individual biases, leading to more robust results than EMCDR and PTUPCDR.
For example, CMF improves NDCG@30 by 2.99\% over EMCDR-MF in the Game and 2.14\% in Video.
There might be a reason for this since EMCDR and PTUPCDR focus on individual user representations, without taking into consideration correlations between users.
In DPMCDR, we hierarchically model this underlying preference commonality with probabilistic domain-level and cross-domain preferences.

\subsubsection*{Variational Information Bottleneck improves Generalization}
Based on information theory, DisenCDR and CRDIB are state-of-the-art CDR approaches tailored to sift out irrelevant information for effective transfer, with impressive performance gaps over others.
Similarly, DPMCDR also includes VIB in the predictive objectives for stronger generalization.
Moreover, DPMCDR extends VIB to the optimization of latent domain-level representations, which further improves expressivity.

\subsubsection*{Distributional Implicit Matching outperforms Explicit Mapping}
Again, NOCDR does not offer reliable explicit correspondence in the two domains.
Unfortunately, this renders the inability of previous approaches relying on deterministic {\it explicit} mapping.
Instead, DPMCDR aligns the cross-domain invariant preferences of both domains derived by random groups of users, with {\it distributional preference matching}.
It does not require and never imposes the alignment on an individual basis.
Therefore, DPMCDR consistently and significantly outperforms state-of-the-art in all metrics with NOCDR.
\begin{table}
    \caption{partial-overlapped CDR scenarios.}
    \label{tab: partial overlap}
    \centering
    \resizebox{\linewidth}{!}{%
    \begin{tabular}{l|ccc|ccc}
    \hline
    \multirow{2}{*}{\textbf{Methods}} & \multicolumn{3}{c|}{Cellphone}                      & \multicolumn{3}{c}{Electronic}                        \\ \cline{2-7} 
                                      & MRR             & NDCG@30         & HR@30           & MRR              & NDCG@30          & HR@30           \\ \hline
    \textbf{DisenCDR}                 & 0.0619          & 0.0799          & 0.1743          & 0.0726           & 0.0956           & 0.2083          \\
    \textbf{CDRIB}                    & {\ul 0.0726}    & {\ul 0.1130}    & {\ul 0.2880}    & {\ul 0.0950}     & {\ul 0.1381}     & {\ul 0.3246}    \\
    \textbf{DPMCDR}                   & \textbf{0.0777} & \textbf{0.1195} & \textbf{0.3020} & \textbf{0.1080}  & \textbf{0.1548}  & \textbf{0.3550} \\ \hline
    Improve(\%)                       & \textbf{7.03\%} & \textbf{5.70\%} & \textbf{4.87\%} & \textbf{13.68\%} & \textbf{12.10\%} & \textbf{9.35\%} \\
    \hline
    \end{tabular}
    }

    \resizebox{\linewidth}{!}{%
    \begin{tabular}{l|ccc|ccc}
    \hline
    \multirow{2}{*}{\textbf{Methods}} & \multicolumn{3}{c|}{Cloth}                          & \multicolumn{3}{c}{Sport}                             \\ \cline{2-7} 
                                      & MRR             & NDCG@30         & HR@30           & MRR              & NDCG@30          & HR@30           \\ \hline
    \textbf{DisenCDR}                 & 0.0224          & 0.0344          & 0.1222          & 0.0256           & 0.0495           & {\ul 0.1740}    \\
    \textbf{CDRIB}                    & {\ul 0.0464}    & {\ul 0.0702}    & {\ul 0.1825}    & {\ul 0.0331}     & {\ul 0.0528}     & 0.1485          \\
    \textbf{DPMCDR}                   & \textbf{0.0468} & \textbf{0.0713} & \textbf{0.1847} & \textbf{0.0396}  & \textbf{0.0637}  & \textbf{0.1761} \\ \hline
    Improve(\%)                       & \textbf{0.97\%} & \textbf{1.58\%} & \textbf{1.22\%} & \textbf{19.61\%} & \textbf{20.74\%} & \textbf{1.20\%} \\
    \hline
    \end{tabular}
    }
\end{table}

\subsection{Overlapped Scenario (RQ 2)}
Having demonstrated its performance in NOCDR, we now investigate whether DPMCDR can handle OCDR with comparable ability.
During training, we consider a partial-overlapping setting in which 85\% of the users do not overlap between domains and 15\% do.
We examine how well DPMCDR performs against consistent second-best performers in NOCDR, i.e., CDRIB and DisenCDR, under two cross-domain scenarios (Cellphone-Electronic and Cloth-sport) in Table (\ref{tab: partial overlap}).

Note that CDRIB and DisenCDR are methods designed for OCDR with bespoke components that leverage explicit shared information between domains, whereas DPMCDR does not.
However, we find that DPMCDR, which treats all users equally regardless of overlap, consistently outperforms CDRIB and DisenCDR.
For example, DPMCDR improves MRR by 7.03\% in Cellphone and 13.68\% in Electronic, demonstrating superior performance.
Our empirical results suggest that DPMCDR is effective in capturing cross-domain knowledge and generating accurate recommendations for both OCDR and NOCDR.

\subsection{Ablation Studies (RQ 3)}
We have discussed the four highlighted designs in DPMCDR in terms of overall performance.
We now present ablation studies to assess the utility of each.
Specifically,
(A) contains {\it Deterministic Graph Encoder} only, with all other components are excluded.
(B) further introduces {\it user-level Stochastic Latent Preference Identifier}, i.e., $q(\mathbf{z}_{1}| \tilde{\mathbf{h}})$, and {\it user-specific optimzer}.
(C) employ complete {\it Stochastic Latent Preference Identifier} and {\it Distributional Preference Matching}, but without two VIB-based predictive {\it user-specific optimizers} and {\it domain-specific optimizers}.
(D) keeps all components of DPMCDR except for {\it domain-specific optimizers}.
Here we only showcase the results for Game-Video due to page limitations.

\begin{figure}
    \centering
    \begin{subfigure}[b]{.45\linewidth}
        \centering
        \includegraphics[width=\textwidth]{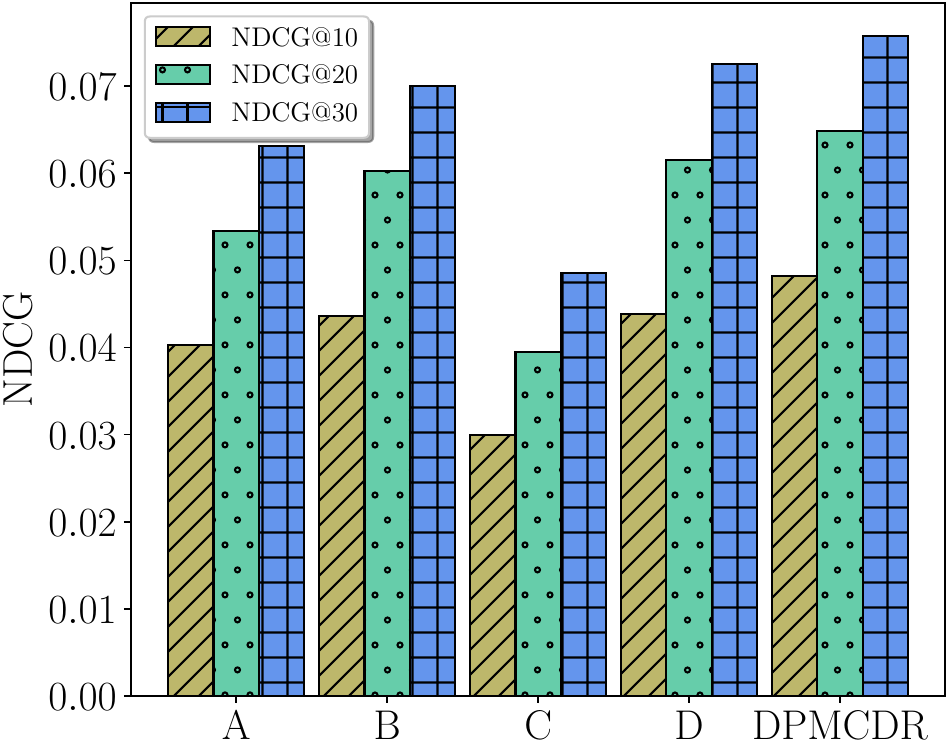}
        \label{fig:game_ndcg}
    \end{subfigure}
    \begin{subfigure}[b]{.45\linewidth}
        \centering
        \includegraphics[width=\textwidth]{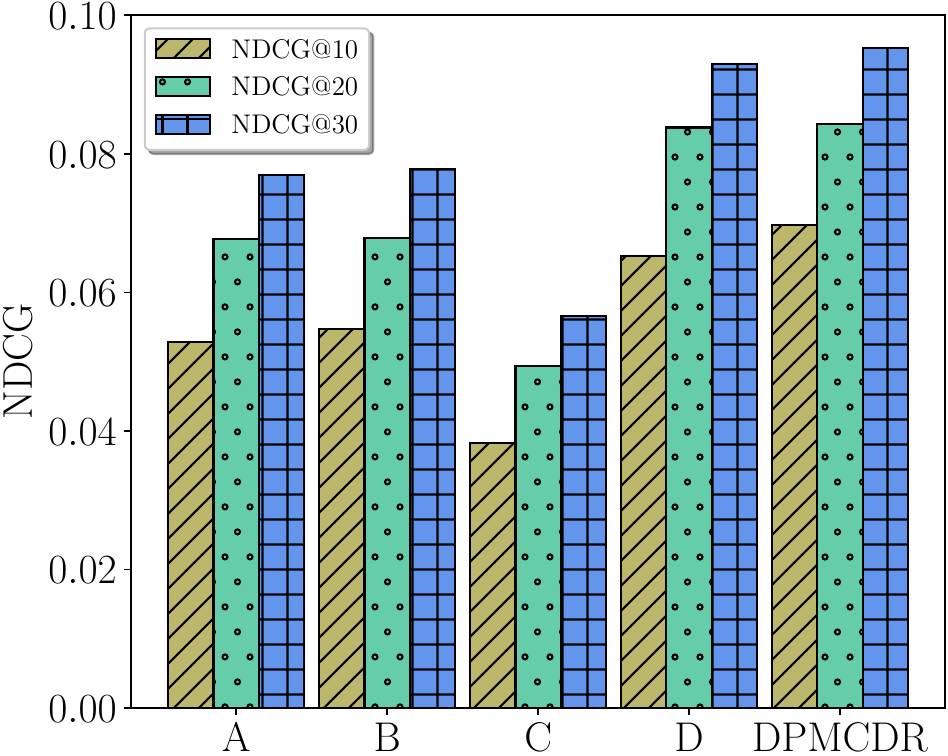}
        \label{fig:video_ndcg}
    \end{subfigure}
    \begin{subfigure}[b]{.45\linewidth}
        \centering
        \includegraphics[width=\textwidth]{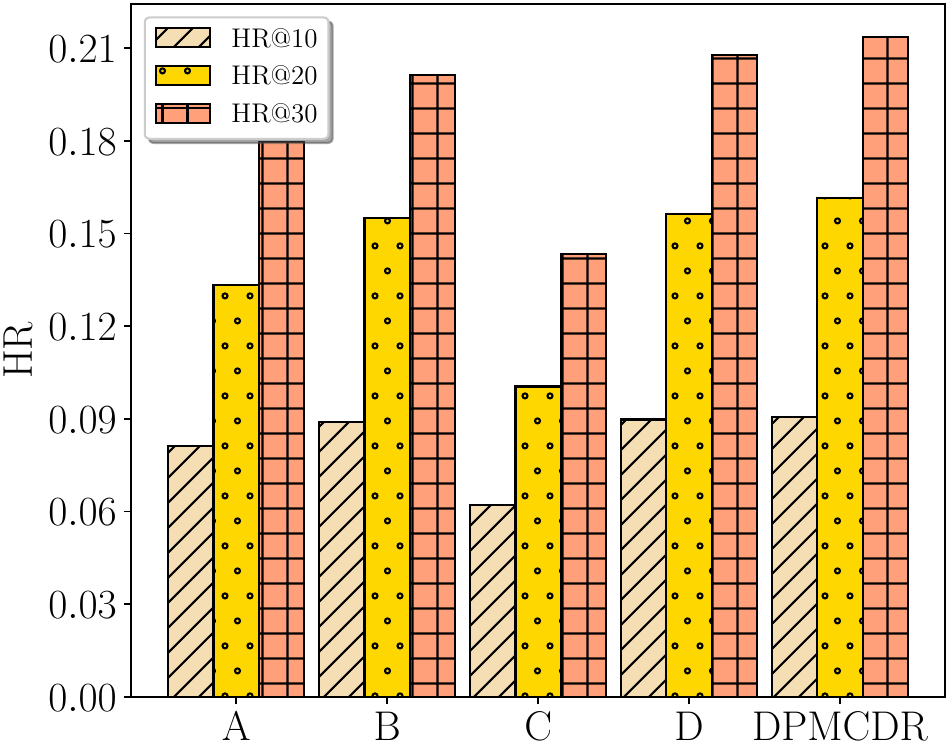}
        \label{fig:game_hr}
    \end{subfigure}
    \begin{subfigure}[b]{.45\linewidth}
        \centering
        \includegraphics[width=\textwidth]{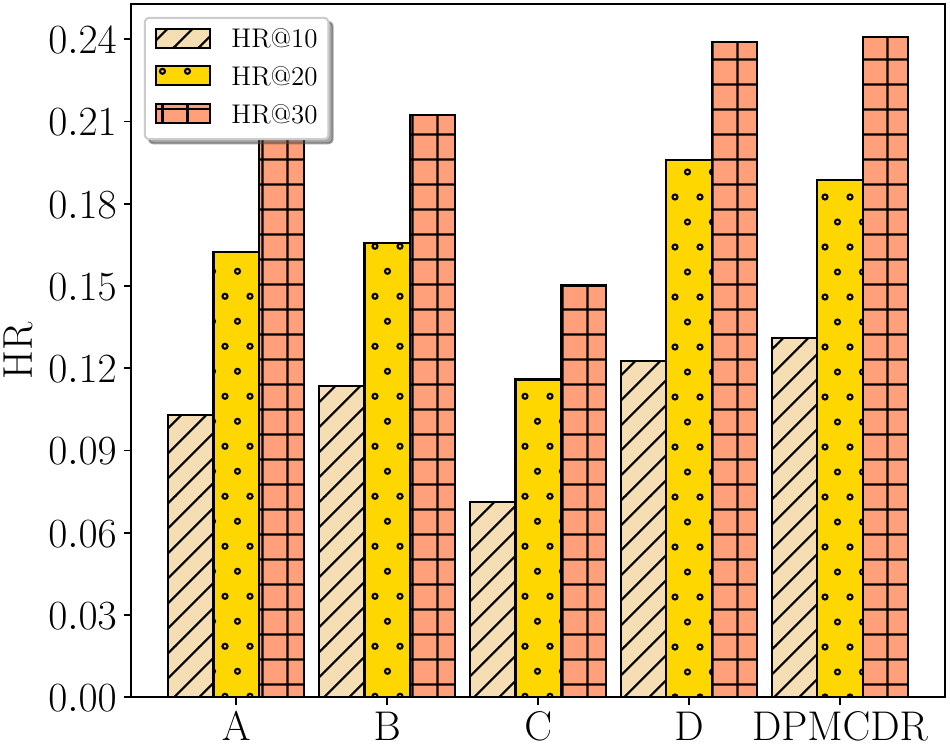}
        \label{fig:video_hr}
    \end{subfigure}
    \vskip -0.05in
    \caption{Ablation Study in the Game(left)-Video(right) scenario.}
    \label{fig: ndcg_hr}
\end{figure}

Fig.~(\ref{fig: ndcg_hr}) compares the NDCG and HR of four variants and full DPMCDR. 
Variant (A) reports 6.32\% in NDCG@30 and 17.96\% in HR@30, outperforming most CDR models (except for CDRIB) and all SDR baselines.
Including {\it user-specific Stochastic Latent Preference Identifier} (variant B) would give rise to performances.
The worst performer of all variants is variant (C) since no predictive objectives are involved.
The domain-level preference cannot be reasonably derived without ground-truth observations.
In comparison, variant (D) injects additional {\it user-specific optimizers}, considerably improving performance over variant (C).
With all modules equipped, DPMCDR~(full) achieves the best performance on all metrics.

\begin{figure}
    \centering
    \begin{subfigure}[b]{.24\linewidth}
        \centering
        \includegraphics[width=\textwidth]{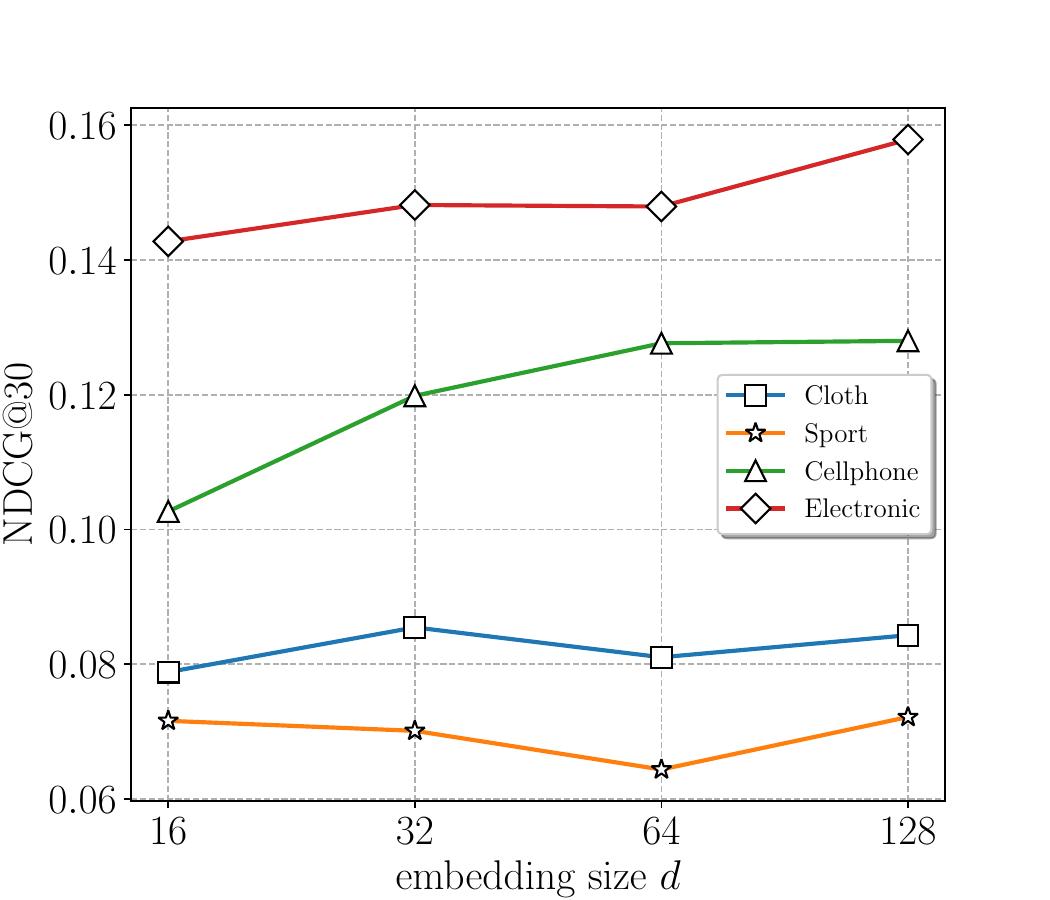}
    \end{subfigure}
    \begin{subfigure}[b]{.24\linewidth}
        \centering
        \includegraphics[width=\textwidth]{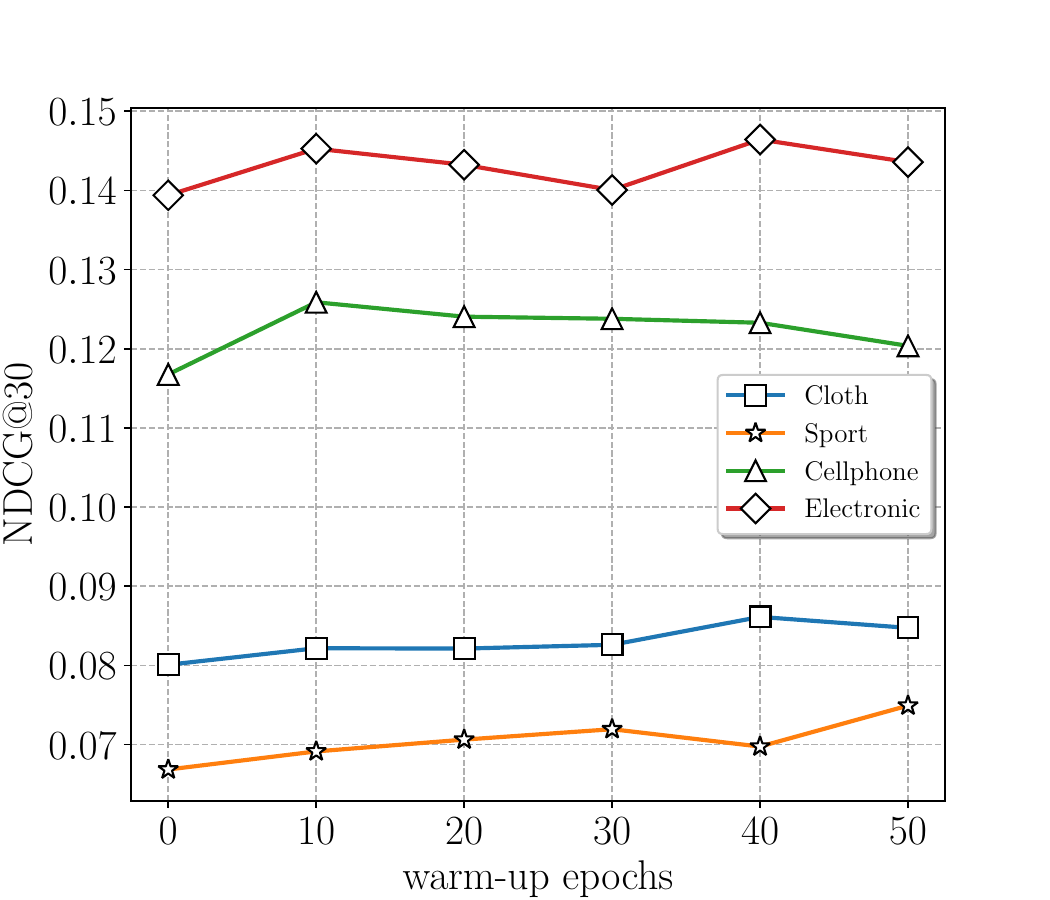}
    \end{subfigure}
    \begin{subfigure}[b]{.24\linewidth}
        \centering
        \includegraphics[width=\textwidth]{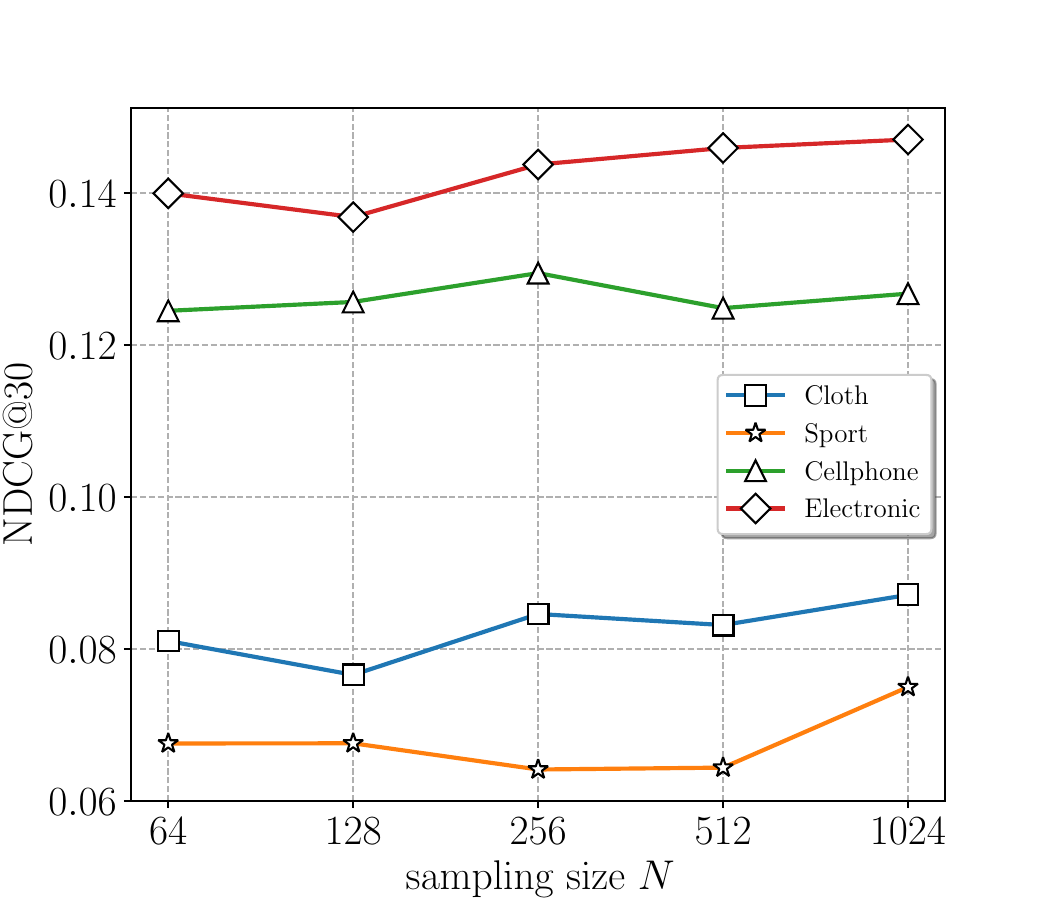}
    \end{subfigure}
    \begin{subfigure}[b]{.24\linewidth}
        \centering
        \includegraphics[width=\textwidth]{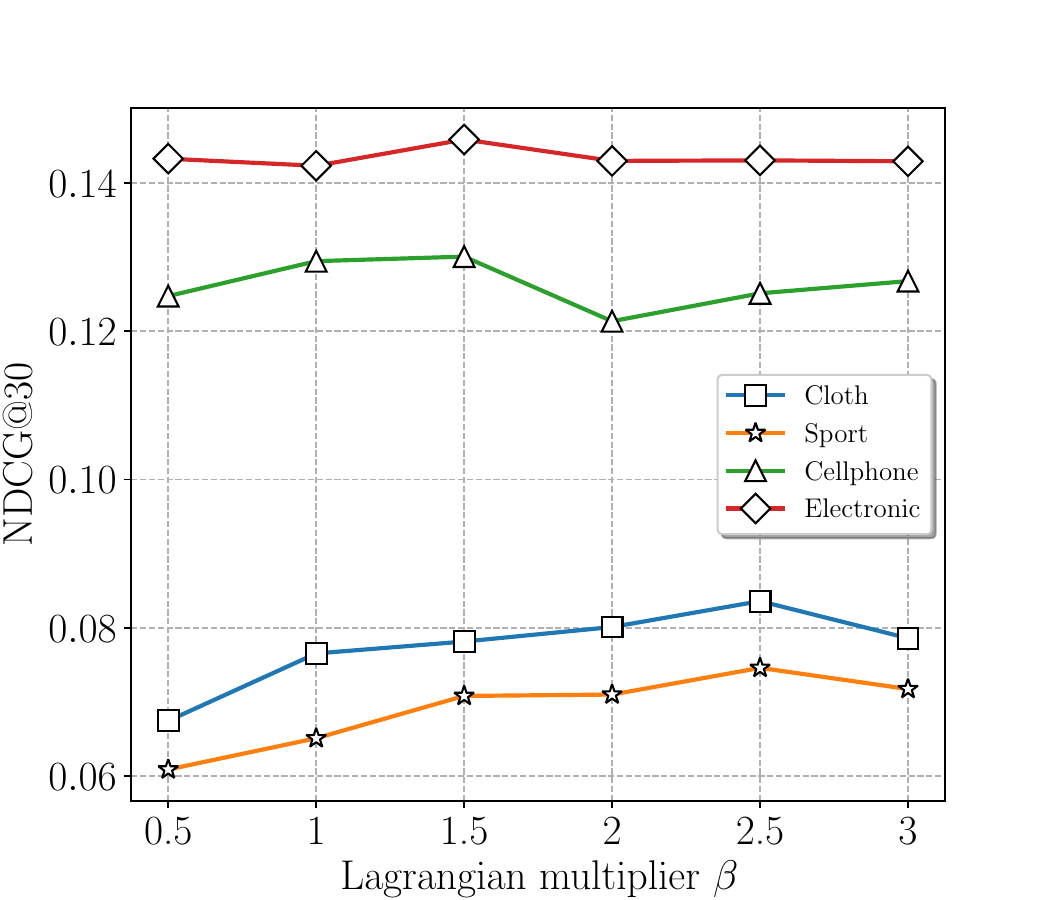}
    \end{subfigure}
    \\
    \begin{subfigure}[b]{.24\linewidth}
        \centering
        \includegraphics[width=\textwidth]{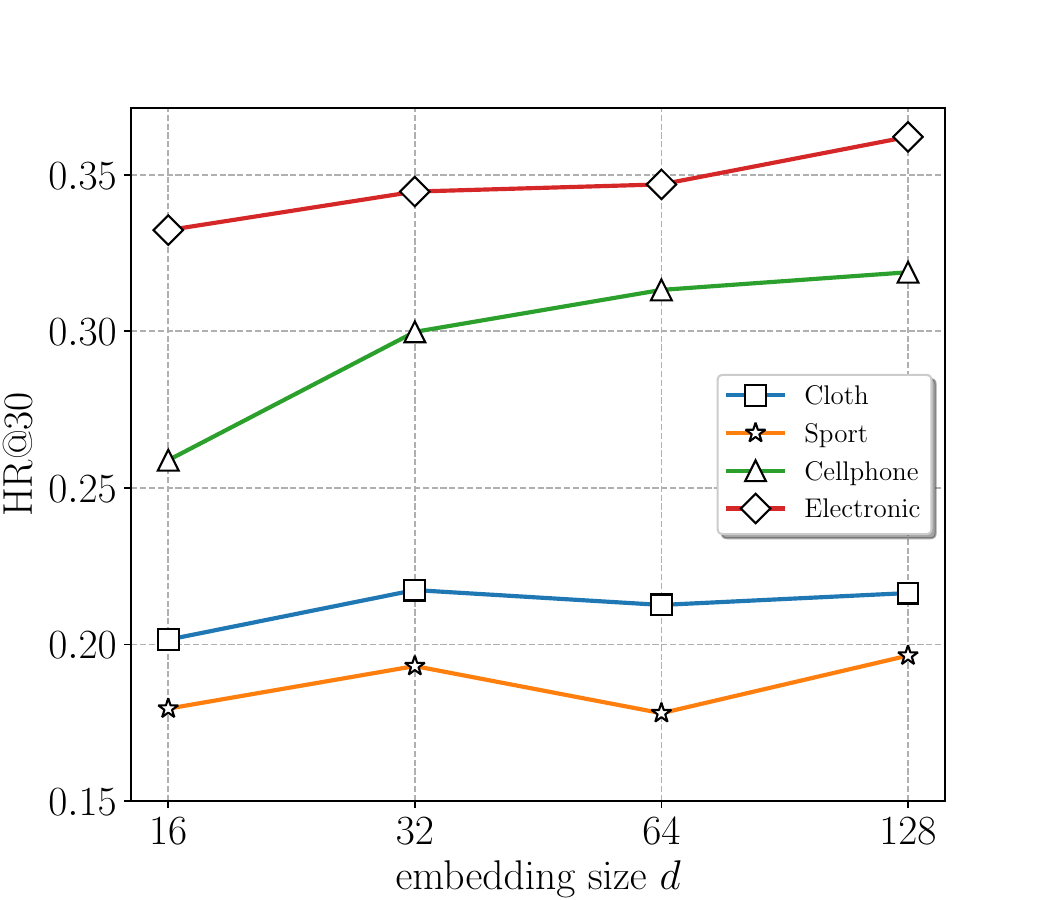}
    \end{subfigure}
    \begin{subfigure}[b]{.24\linewidth}
        \centering
        \includegraphics[width=\textwidth]{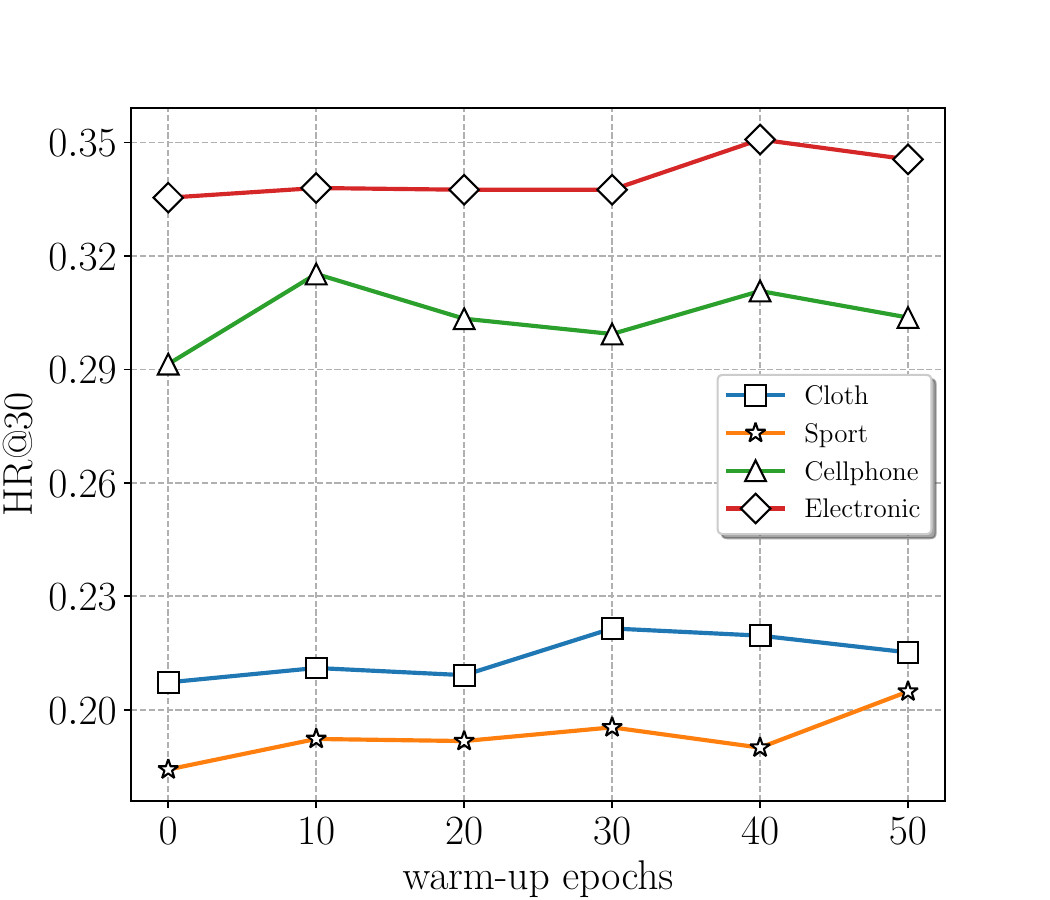}
    \end{subfigure}
    \begin{subfigure}[b]{.24\linewidth}
        \centering
        \includegraphics[width=\textwidth]{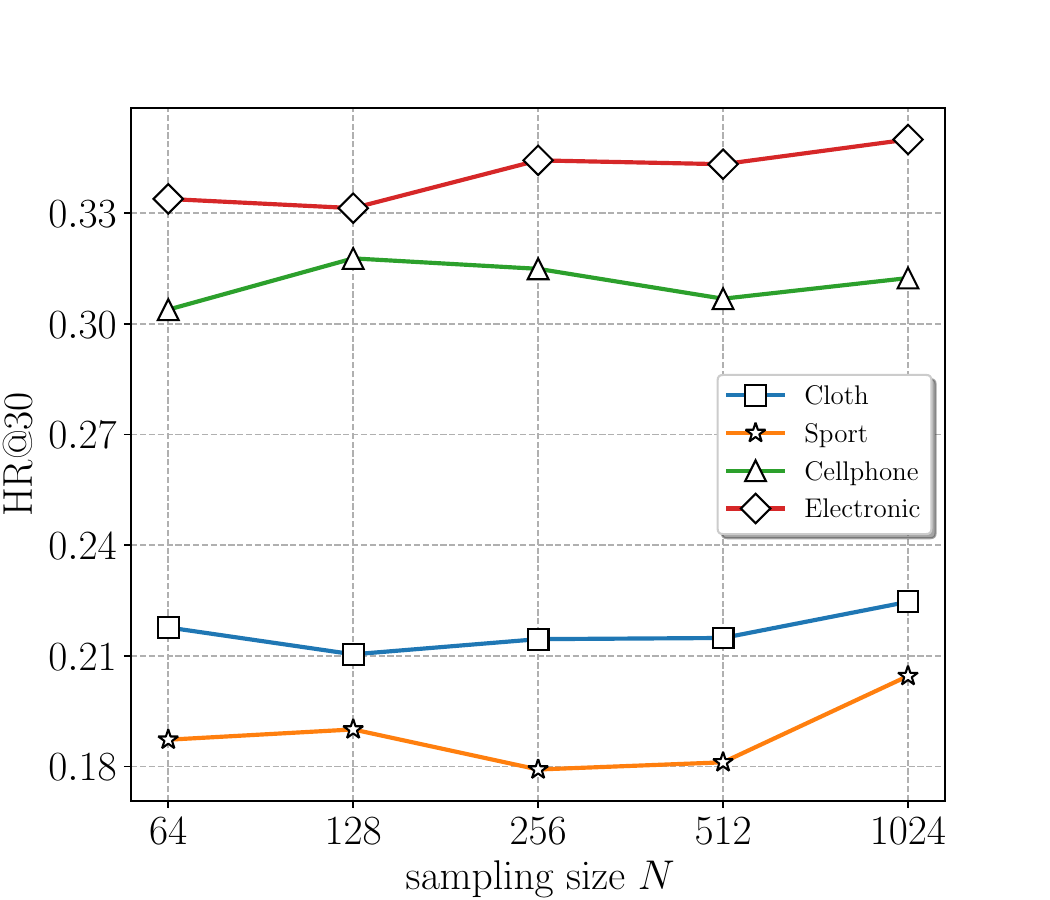}
    \end{subfigure}
    \begin{subfigure}[b]{.24\linewidth}
        \centering
        \includegraphics[width=\textwidth]{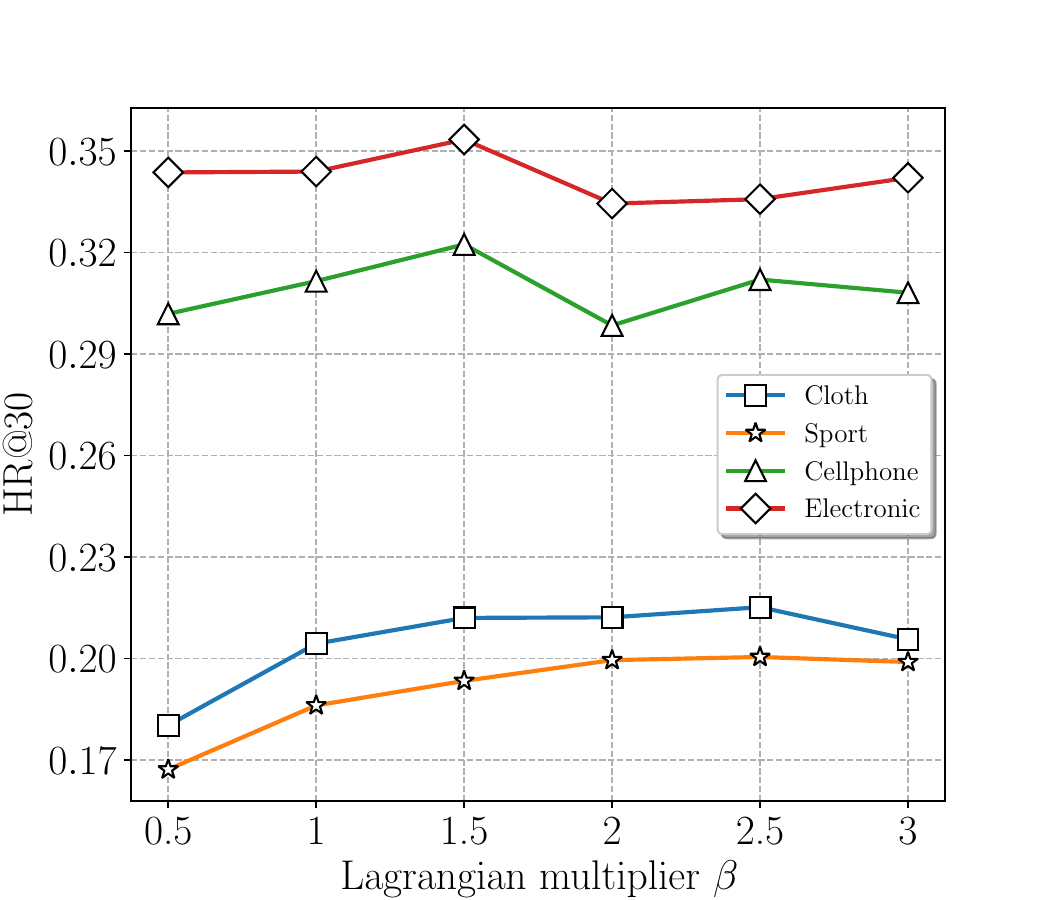}
    \end{subfigure}
    \caption{Parameter Sensitivity}
    \label{fig: embed}
\end{figure}

\subsection{Parameter sensitivity (RQ 3)}
We evaluate four important hyperparameters, i.e., embedding size $d$, warm-up epoch, sampling size $N$, and Lagrangian multiplier $\beta$, to see how varying these values would affect the performance of DPMCDR.
We only show the results of Cloth-Sport and Cellphone-Electronic given limited pages.

As shown in Fig.~(\ref{fig: embed}), we search for the best embedding size within \{16, 32, 64, 128\}.
In the Cellphone-Electronic scenario, all metrics increase with embedding size $d$ and reach an optimum when $d$=128.
Concerning Cloth-Sport, NDCG and HR show fluctuations but perform best when $d$=128.
It is still only marginally better than $d$=32.
Increasing $d$ may improve performance but at the cost of greater computational costs.

Warm-up epochs refer to the period before {\it distributional preference matching} joins model training.
Optimizing for preference matching can be detrimental if latent representations are not strong enough.
We search the best warm-up epochs from \{0, 10, 20, 30, 40, 50\}.
We find that Different tasks lead to different peak performance epochs.
Nevertheless, the model retains a higher performance after warming up than without, underlining the need for cross-domain matching.

As for sampling size of groups $N$ in {\it Stochastic Latent Preference Identifier}, we vary $N$ within \{64, 128, 256, 512, 1024\}.
Increasing $N$ from 64 to 256 leads to fluctuations in HR@30 and NDCG30. 
However, we observe an improvement in most results if we enlarge the sampling size to 1024.
Larger sampling sizes enhance DPMCDR to aggregate more users, enabling accurate derivation of domain preference distribution.

We lastly evaluate Lagrangian multiplier $\beta$ ranging from 0.5 to 3 with a step length of 0.5.
The choice of $\beta$ depends on the specific CDR task.
While $\beta$=2.5 gives the best results for Cloth-Sport CDR prediction, this value hampers the performance for the Cellphone-Electronic task.
Although best values differ across CDR tasks, both domains follow similar trends.

\section{Conclusion and Future Works}
We have proposed a distributional cross-domain invariant preference matching approach, DPMCDR, based on a shared latent space that aligns cross-domain invariant preferences for Cross-Domain Recommendation.
We presumed deterministic user representations as observations from the continuous preference prior distribution and approximated its posterior with random groups of users, drawing latent user-wise correlations and identifying commonality within latent representations.
The latent representations are further grounded in a shared latent space to match the predictive distributions of cross-domain invariant preferences described by two domains.
Our optimization further improved the cross-domain generalization of invariant preferences under the variational information bottleneck principle.
Extensive experiments demonstrated that DPMCDR consistently outperforms state-of-the-art with a range of metrics, regardless of the existence of overlap users.
The advantages of DPMCDR are particularly highlighted in NOCDR where no overlap information can be used.
This case still allows DPMCDR to capture user preference commonality and address the limitation of existing studies.
Future plans include extending distributional matching to the multi-domain settings, where varied divergences among domains must be addressed to avoid negative transfer.
Adapting such domain-wise similarities into modeling might be an interesting workaround.

\bibliographystyle{IEEEtran}
\bibliography{reference}

\end{document}